\documentclass[reprint,amsmath,amssymb,aps,prl]{revtex4-1}

\usepackage{graphicx}
\usepackage{dcolumn}
\usepackage{bm}
\usepackage{physics}

\usepackage{hyperref}

\usepackage{enumerate}  
\usepackage{float}
\usepackage{xargs}  
\usepackage{setspace}
\usepackage[export]{adjustbox}
\usepackage{fontawesome}

\begin{document}

\title{Strong optical coupling through superfluid Brillouin lasing}

\author{Xin He$^*$}
\affiliation{ARC Centre of Excellence for Engineered Quantum Systems, School of Mathematics and Physics, The University of Queensland, St Lucia, Australia. \\$^*$These authors contributed equally. $^\dagger$To whom correspondence should be addressed, c.baker3@uq.edu.au}
\author{Glen I. Harris$^*$}
\affiliation{ARC Centre of Excellence for Engineered Quantum Systems, School of Mathematics and Physics, The University of Queensland, St Lucia, Australia. \\$^*$These authors contributed equally. $^\dagger$To whom correspondence should be addressed, c.baker3@uq.edu.au}
\author{Christopher G. Baker$^{*,\dagger}$}
\affiliation{ARC Centre of Excellence for Engineered Quantum Systems, School of Mathematics and Physics, The University of Queensland, St Lucia, Australia. \\$^*$These authors contributed equally. $^\dagger$To whom correspondence should be addressed, c.baker3@uq.edu.au}
\author{Andreas Sawadsky}
\affiliation{ARC Centre of Excellence for Engineered Quantum Systems, School of Mathematics and Physics, The University of Queensland, St Lucia, Australia. \\$^*$These authors contributed equally. $^\dagger$To whom correspondence should be addressed, c.baker3@uq.edu.au}
\author{Yasmine L. Sfendla}
\affiliation{ARC Centre of Excellence for Engineered Quantum Systems, School of Mathematics and Physics, The University of Queensland, St Lucia, Australia. \\$^*$These authors contributed equally. $^\dagger$To whom correspondence should be addressed, c.baker3@uq.edu.au}
\author{Yauhen P. Sachkou}
\affiliation{ARC Centre of Excellence for Engineered Quantum Systems, School of Mathematics and Physics, The University of Queensland, St Lucia, Australia. \\$^*$These authors contributed equally. $^\dagger$To whom correspondence should be addressed, c.baker3@uq.edu.au}
\author{Stefan Forstner}
\affiliation{ARC Centre of Excellence for Engineered Quantum Systems, School of Mathematics and Physics, The University of Queensland, St Lucia, Australia. \\$^*$These authors contributed equally. $^\dagger$To whom correspondence should be addressed, c.baker3@uq.edu.au}
\author{Warwick P. Bowen}
\affiliation{ARC Centre of Excellence for Engineered Quantum Systems, School of Mathematics and Physics, The University of Queensland, St Lucia, Australia. \\$^*$These authors contributed equally. $^\dagger$To whom correspondence should be addressed, c.baker3@uq.edu.au}

\date{\today}


\maketitle

{\bf
Brillouin scattering has applications ranging from signal processing~\cite{Kittlaus_NatComm_17,MicrowaveSynthesizer}, sensing~\cite{Eggleton_Review} and microscopy~\cite{scarcelli2008confocal}, to quantum information~\cite{Renninger_NatPhys_18} and fundamental science~\cite{HighFrequency, kashkanova2017superfluid}. Most of these applications rely on the electrostrictive interaction between light and phonons~\cite{Eggleton_Review,kashkanova2017superfluid, giorgini_stimulated_2018}. Here we show that in liquids optically-induced surface deformations can provide an alternative and far stronger interaction. This allows the demonstration of ultralow threshold Brillouin lasing and strong phonon-mediated optical coupling for the first time. This form of strong coupling is a key capability for Brillouin-reconfigurable optical switches and circuits~\cite{Ruesink_NatComms_18,Shen_NatComms_18}, for photonic quantum interfaces~\cite{SafaviNaeini_NJP_11}, and  to generate synthetic electromagnetic fields~\cite{Fang_NatPhys_17,Schmidt_Optica_15}. While applicable to liquids quite generally, our demonstration uses superfluid helium. Configured as a Brillouin gyroscope~\cite{BrillouinGyroscope} this provides the prospect of measuring superfluid circulation with unprecedented precision, and to explore the rich physics of quantum fluid dynamics, from quantized vorticity to quantum turbulence~\cite{sachkou_coherent_2019,Gauthier1264}.}

Brillouin scattering is an optomechanical process that couples two optical waves via their interaction with travelling acoustic phonons. In the electrostrictive interaction usually employed, the optical electric field induces strain in a bulk medium, and the generated phonons scatter light between the two optical waves via refractive index changes caused by the medium's photoelasticity. However, the inherent weakness of this interaction presents a significant challenge~\cite{Eggleton_Review}, necessitating the use of high optical powers and prohibiting some applications. This can be alleviated by resonant enhancement in an optical cavity, which has allowed recent demonstrations of ultralow linewidth lasers~\cite{ObservationOfBrillouin, Gundavarapu_IntegratedBrillouin}, Brillouin lasing in liquid droplets~\cite{giorgini_stimulated_2018}, non-reciprocal optical transport~\cite{Dong_BIT_reciprocity}, Brillouin gyroscopes~\cite{BrillouinGyroscope} and low-noise microwave oscillators~\cite{MicrowaveSynthesizer}. Alternatively, optically-induced deformations of the boundary of the medium can be leveraged to provide a Brillouin interaction, with scattering induced by the effective refractive index-modulation caused by the deformation. In purpose-engineered solid structures these surface interactions can be made comparable to, or even exceed, the native electrostriction~\cite{Shin_NatCom_13,VanLaer_NatPhot_2015,wolff_stimulated_2015}. 

\begin{figure*}[ht!]
\centering
\includegraphics[width=\textwidth]{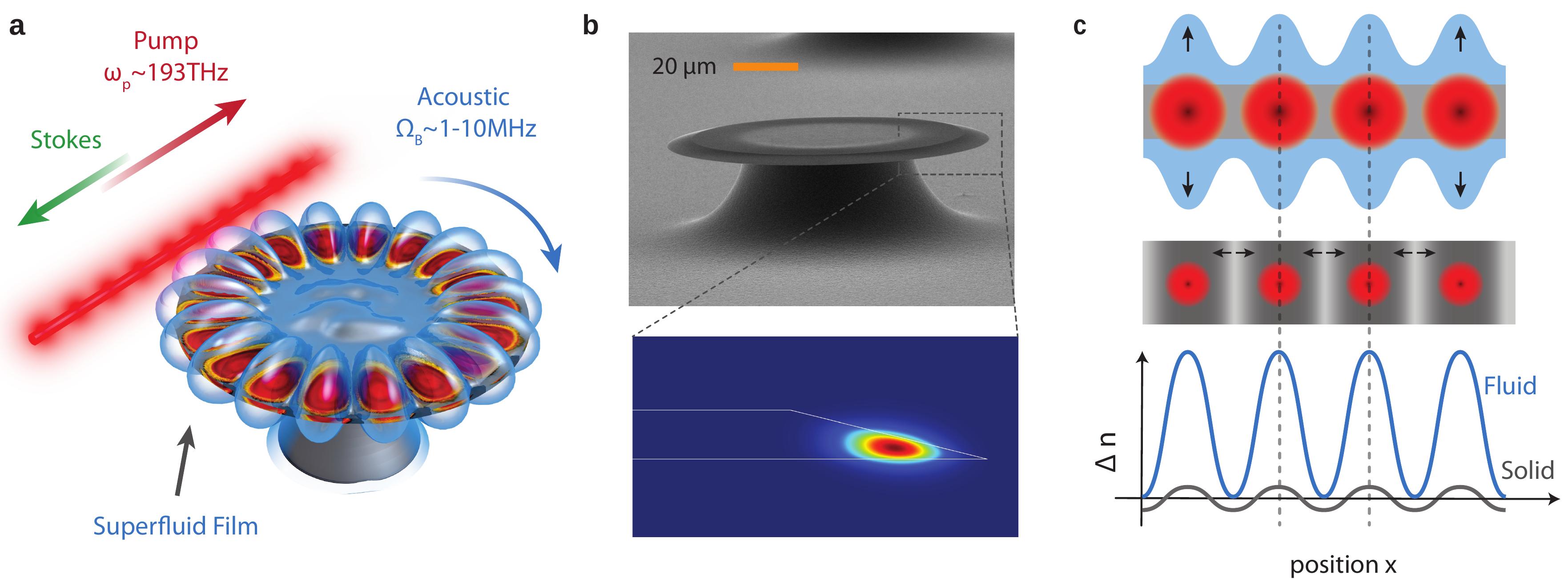}%
\caption{(a) Illustration of the backward Brillouin scattering process in the superfluid optomechanical resonator (b) Top: Scanning Electron Microscope (SEM) image of a silica microdisk resonator on the silicon chip (radius 40 $\mu$m, thickness 2 $\mu$m). Bottom: cross-section showing the electric field norm of a quasi-TE mode supported by the microdisk. (c) Schematic illustration of the Brillouin interaction in our superfluid system (top) and in a solid material where the optical field is responsible for compression and rarefaction of the medium (middle). Due to threefold resonant enhancement, the optically driven refractive index modulation (bottom) is larger in the fluid case. 
}
\label{fig:idea}
\end{figure*}

Here, we transfer the concept of deformation-induced Brillouin scattering to liquid media, specifically a few-nanometer-thick superfluid helium film that coats the surface of a silica microdisk cavity and couples to its whispering gallery modes (see Fig.~\ref{fig:idea}a,b) via perturbation of their evanescent field. Similar to other liquids~\cite{kaminski_ripplon_2016}, the superfluid film has an exceedingly weak restoring force, affording a compliant dielectric interface that easily deforms in the presence of optical forces~\cite{harris_laser_2016, baker_theoretical_2016}, as illustrated in Fig.~\ref{fig:idea}c. This offers the potential for very large surface deformations and consequently extreme interaction strengths. We show that it allows radiation-pressure interactions with acoustic phonons that are over two orders of magnitude stronger than have been achieved using electrostriction in similar-sized microcavities~\cite{ObservationOfSpontaneous, Bahl_NatComm_2013}. These interactions are enhanced by a further order of magnitude by the superfluid fountain pressure, where optical absorption-induced entropy gradients induce superfluid flow~\cite{mcauslan_microphotonic_2016}. 

In addition to confining light, the microdisk used in our experiments provides confinement for travelling whispering gallery-like surface acoustic waves in the superfluid film. These acoustic waves are co-located with the optical whispering gallery modes close to the perimeter of the disk (see Fig.~\ref{fig:idea}a) and are a class of  {\it third-sound}~\cite{ThirdSound}, which manifests as film thickness fluctuations with a restoring force provided by the van der Waals interaction. They  generate an effective refractive index grating that scatters photons between forward and backward propagating optical whispering gallery modes, forming a Brillouin optomechanical system. The thin wedge-shaped perimeter of the microdisk (see Fig.~\ref{fig:idea}b) is engineered to maximize the optical evanescent field at the location of the acoustic wave~\cite{baker_theoretical_2016}. Combined with the co-location of light and acoustic waves, this enhances the achievable radiation pressure coupling by several orders of magnitude compared to conventional optomechanics experiments previously performed with thin superfluid films~\cite{harris_laser_2016, mcauslan_microphotonic_2016, sachkou_coherent_2019}.

In the backward Brillouin scattering process demonstrated here, a pump photon of frequency $\omega_p$ is scattered into a lower frequency counter-propagating Stokes photon at $\omega_s$, and a forward-travelling acoustic phonon of frequency $\Omega_B$ and wavelength $\lambda_B$ in the superfluid, where energy and momentum conservation dictate that $\omega_p=\omega_s+\Omega_B$ and that the optical wavelength  $\lambda_{\mathrm{light}} = 2\lambda_{B}$ (see Fig.~\ref{fig:idea}a).
In solids, the deformation due to electrostriction is determined by the Young's modulus, $E$.
By contrast, in our experiments, regions of high light intensity inside the resonator continuously deform a fluid interface by drawing in more superfluid by means of an optical gradient force and the superfluid fountain effect~\cite{ashkin1986observation, mcauslan_microphotonic_2016}. Here, the equivalent of the Young's modulus is the van der Waals pressure on the surface of the fluid, $P_v = 3 \alpha_v \rho/d^3$, where $\alpha_v$ is the van der Waals coefficient, $\rho$ is the fluid density and $d$ is its thickness. While both electrostriction and surface deformation result in a periodically modulated refractive index grating which scatters pump light (illustrated for our case in Fig. \ref{fig:idea}(c)), for typical few nanometer-thick films, the van der Waals pressure is over six orders of magnitude lower than the Young's modulus of typical Brillouin active solids ($\sim$ kPa versus GPa). As a result, the modulation achieved per photon can be much larger, enabling the ultra-low threshold lasing reported here, and providing the future prospect to reach quantum regimes.
The reduced stiffness greatly reduces the speed of sound $c$, from kilometers per second for solids to meters per second for superfluid films.  This results in a Brillouin shift $\Omega_B/2\pi=2c/\lambda_{\mathrm{light}}$ in the megahertz range, versus gigahertz range in many solids at telecom wavelengths. 

These unique features offer two advantages. First, because of the small frequency shift, the Stokes beam is naturally resonant with the degenerate whispering gallery mode that counter-propagates with the pump mode, alleviating the need for potentially complex spectral engineering of the resonator~\cite{BrillouinGyroscope, ObservationOfBrillouin} and allowing resonant enhancement that is independent of the device size, down to  mode volumes in the range of the optical wavelength cubed.
Quite generally, such miniaturization is desirable since it enhances the strength of the Brillouin interaction, but so far this has been hampered by the inability to meet momentum and energy conservation criteria in micron-scale devices with sparse optical spectra.
Second, and a key result of this work, the strong Brillouin scattering enables the emergence of mechanically-mediated strong optical coupling, hybridising the initially degenerate forward and backward propagating fields.

\begin{figure}[ht!]
\centering
\includegraphics[width=\columnwidth]{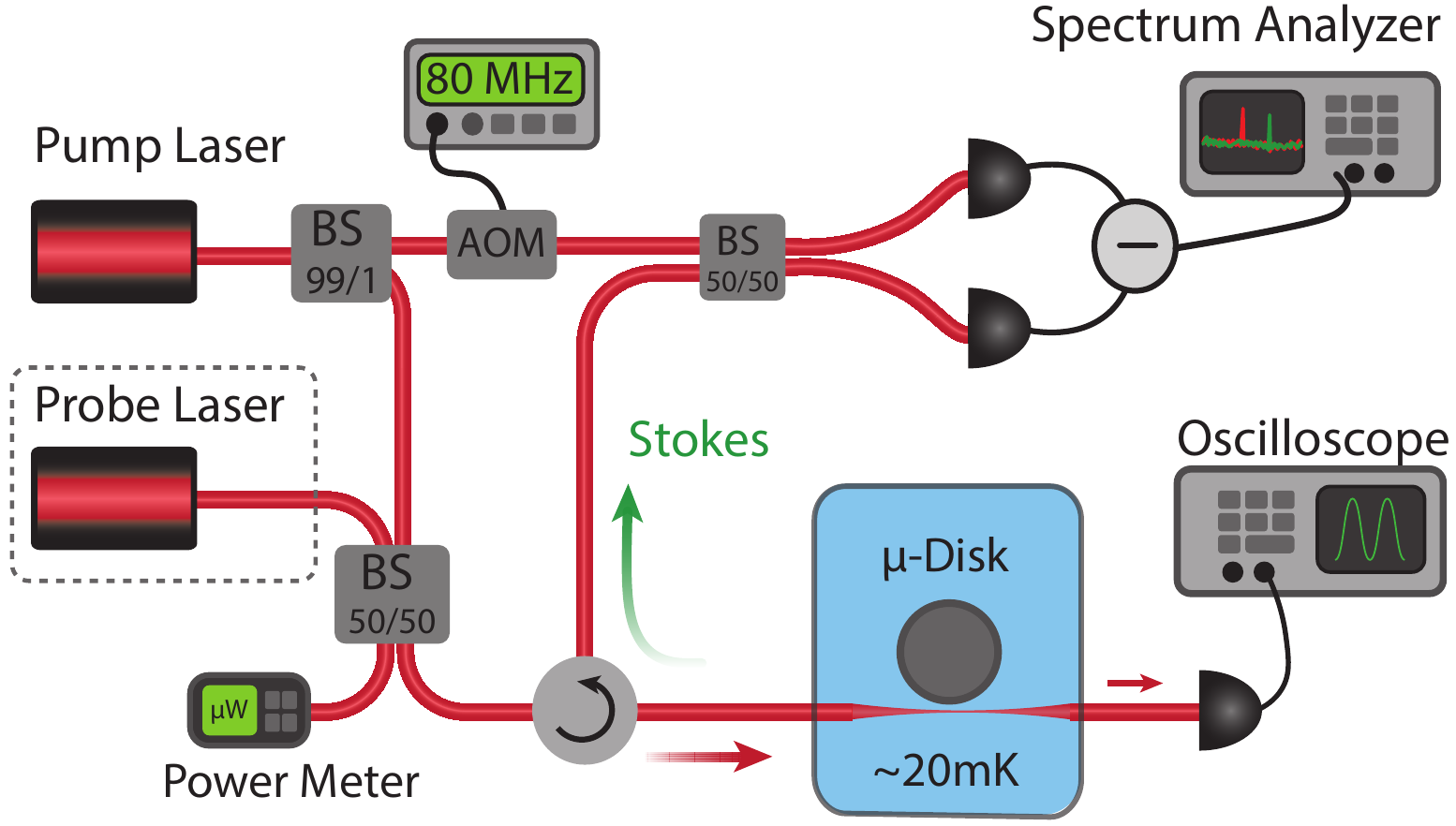}%
\caption{Schematic of the experimental setup. BS: beamsplitter; AOM: acousto-optic modulator; blue box: sample chamber in the dilution refrigerator.}
\label{fig:setup}
\end{figure}

\begin{figure*}[ht!]
\centering
\includegraphics[width=\textwidth]{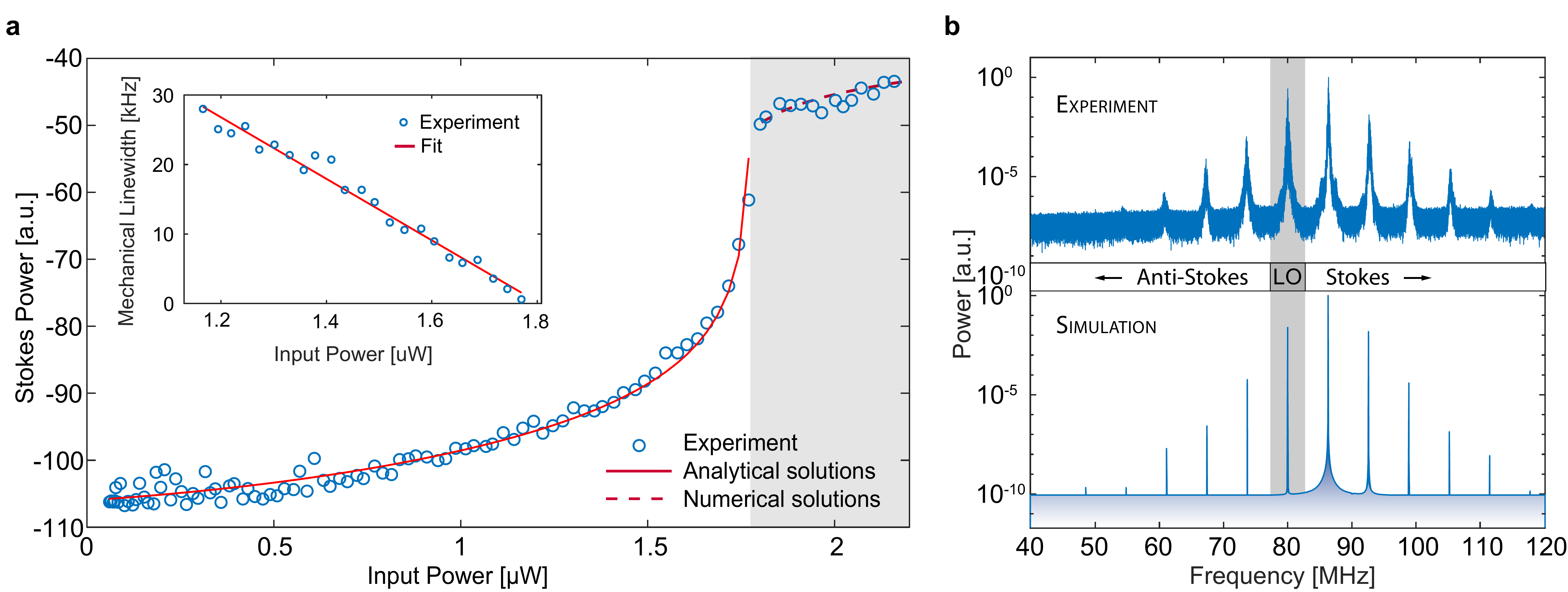}
\caption{(a) Stokes sideband peak power versus input laser power (blue circles). Solid red line is an analytical fit to the data in the non-depleted pump regime; dashed red line is a fit obtained by numerically solving the full equations of motion (see Methods and Supplemental Material). Inset: effective acoustic linewidth $\Gamma_{\mathrm{eff}}$ vs input power, along with fit by Eq.(\ref{eq:Gammaeff2}). Shaded region above 1.8 $\mu$W marks the onset of Brillouin lasing. (b) Brillouin lasing spectra measured at 5.6 $\mu$W input power, (top: experimental spectrum; bottom: numerical solution to theory). LO: frequency shift between pump and heterodyne local oscillator. Note that Stokes sidebands appear at higher frequencies due to the heterodyne detection.}
\label{fig:powersweep}
\end{figure*}

The optical pumping of phonons into the forward traveling acoustic wave in backward Brillouin scattering amplifies the  acoustic wave, modifying its linewidth to
\begin{equation}
\Gamma_{\mathrm{eff}} =\Gamma - \frac{4\, g_{0,\mathrm{tot}}\, g_{0,\mathrm{rp}}}{\kappa}\,\zeta\, n_{\mathrm{cav}},
\label{eq:Gammaeff2}
\end{equation}
where $\Gamma$ is the intrinsic linewidth, $n_{\mathrm{cav}}$ is the intra-cavity photon number, $\kappa$ is the loaded optical linewidth, and $g_{0,\mathrm{tot}}=g_{0,\mathrm{rp}}+g_{0,\mathrm{fp}}$ is the single photon optomechanical coupling rate which includes radiation pressure ($g_{0,\mathrm{rp}}$) and fountain pressure ($g_{0,\mathrm{fp}}$) components (see Supplemental Material). The parameter $\zeta< 1$ quantifies a reduction in Brillouin amplification from the presence of optical backscattering, caused by geometric imperfections of the microdisk. This reduction is due both to the backscattering-induced depletion of the pump field, and to active cooling of the acoustic mode by the backscattered photons, which directly competes with the amplification from the pump. Once the effective linewidth crosses zero, the acoustic damping is offset by the Brillouin amplification. In this regime, spontaneously occurring thermal excitations in the superfluid film are exponentially amplified to a large coherent amplitude, i.e. the system exhibits phonon lasing.

The apparatus used to experimentally probe superfluid thin-film Brillouin lasing is shown in Fig. \ref{fig:setup}, with details given in the Methods. Figure~\ref{fig:powersweep}(a) shows the observed Stokes signal peak power as a function of  input laser power,  while the inset shows the Stokes linewidth versus power. As can be seen, the Stokes wave is amplified with increasing input power in good agreement with theory, and crosses the threshold to lasing at a power of 1.8~$\mu$W. Fitting the linewidth data to Eq.~(\ref{eq:Gammaeff2}) yields an intrinsic acoustic linewidth of $\Gamma/2\pi=85\pm 6$~kHz. We extract a total single photon coupling rate of $g_{0, \mathrm{tot}}/2\pi=133$~kHz, independently of $\zeta$, through numerical simulations that include all dynamic couplings between optical and acoustic modes induced by backscattering. The radiation pressure component is estimated to be $g_{0,\mathrm{rp}}/2\pi=11$ kHz from finite element modeling (see Supplemental Material), leaving a fountain pressure contribution of $g_{0,\mathrm{fp}}/2\pi=122$ kHz. The total single photon coupling rate is more than three orders of magnitude larger than what is achieved using the electrostrictive interaction in similar-sized silica microspheres~\cite{ObservationOfSpontaneous}. Furthermore, it represents a several order-of-magnitude increase compared to previous superfluid optomechanics experiments with microtoroidal cavities~\cite{harris_laser_2016, mcauslan_microphotonic_2016, sachkou_coherent_2019}.
\begin{figure}[t]
\centering
\includegraphics[width=.9\columnwidth]{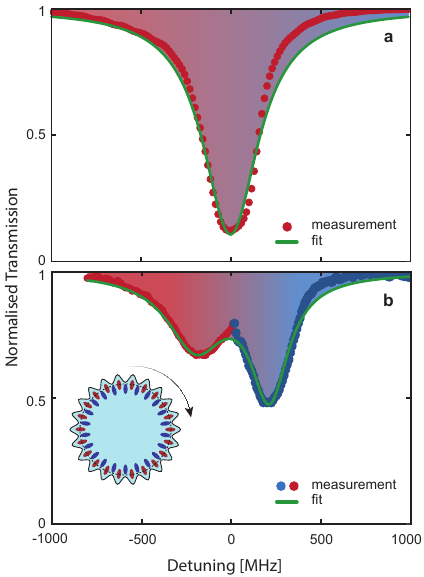}%
\caption{(a) With the pump laser off, the optical mode measured by the probe beam is non-split. (b) At 5.1 $\mu$W pump power, the lasing of Brillouin wave splits the optical mode into a doublet.}%
\label{fig:OpticalSplitting}%
\end{figure}
The fountain pressure is introduced by optical heating through the temperature dependence of the entropy of the film. Gradients in entropy induce a flow of the superfluid component in an analogous manner to optical gradient forces.  Despite the observed strength of this interaction, we verify that it comes in concert with only minimal levels of bulk heating, and with feasible modifications could allow ground state cooling~\cite{Metzger_Nat_04,Jourdan_PRL_08} (see Supplemental Material). 

Above the lasing threshold the pump light is depleted by scattering off the strong superfluid travelling refractive index grating, such that the linearized theory is no longer valid. Furthermore, the presence of optical backscattering introduces a dynamic coupling between optical modes. Therefore, the above-threshold behavior is fitted by numerically solving the full equations of motion of the optical and acoustic fields (see Supplemental Material). These simulations are in good agreement with both the measured power scaling (dashed red curve in Fig. \ref{fig:powersweep}(a)) and the optical power spectra (Fig. \ref{fig:powersweep}(b)), providing a confirmation of the saturation behavior through pump depletion.

In the Brillouin interaction, the lasing of the co-propagating acoustic wave creates  an effective refractive index grating travelling at the speed of sound. 
If the backscattering generated from this grating is sufficiently strong, it is possible to enter a regime of mechanically-mediated photon-photon strong coupling, where photons cycle between the pump and Stokes resonances at a rate faster than the decay rate of the optical fields. This should be distinguished from the {\it phonon-photon} strong coupling previously demonstrated in both conventional~\cite{Verhagen_Nat_2012} and Brillouin~\cite{ObservationOfBrillouin} optomechanics. The signature of this strong coupling is that the pump and Stokes resonances hybridize into a new pair of frequency-shifted optical eigenmodes, with frequencies given by:
\begin{equation}
\omega_\pm = \bar \omega \pm \sqrt{\Omega_B^2/4 + g_{\mathrm{opt}}^2},
\label{eq:splitting}
\end{equation}
where $\bar \omega = (\omega_p + \omega_{s})/2$ is the average of the two bare optical modes that support the pump and Stokes photons, and $g_{\mathrm{opt}}=g_{\mathrm{0,rp}} \, \beta$ is the \emph{phonon-boosted} optomechanical coupling rate with $\beta$ the amplitude of the co-propagating acoustic wave (see Supplemental Material). 

Our superfluid thin-film optomechanical Brillouin laser is naturally suited to observe the mode hybridisation induced by strong coupling due to its combination of high radiation pressure coupling rate $g_{0,\mathrm{rp}}$ and low Brillouin frequency $\Omega_B$. This allows access to the regime where $g_{\rm opt} \gg \Omega_B$ in which the optical modes are shifted by frequencies of $\pm\left[(\Omega_B^2/4 + g_{\mathrm{opt}}^2)^{1/2} -\Omega_B/2\right] \sim \pm g_{\mathrm{opt}}$. By comparison, in the usual regime where the Brillouin frequency is high enough that $\Omega_B \gg g_{\rm opt}$, the optical mode shifts are suppressed by a factor of $\Omega_B/g_{\rm opt}$. 

To spectroscopically explore the regime of strong coupling we implement a pump-probe measurement scheme, shown in Fig.~\ref{fig:setup} and described in the Methods. Figure~\ref{fig:OpticalSplitting} shows the optical resonance measured with this setup. With the pump laser off (Fig. \ref{fig:OpticalSplitting}(a)), no splitting is evident between the clockwise (pump) and counter-clockwise (Stokes) optical modes. Injecting 5~$\mu$W of pump light initiates spontaneous Brillouin lasing, as verified through the reflected light power spectrum.
Due to the high contrast refractive index grating generated by the Brillouin lasing process, two distinct non-degenerate cavity resonances can now be resolved (Fig. \ref{fig:OpticalSplitting}(b)): a lower frequency resonance (red) sensing the peaks of the superfluid  Brillouin wave, and a higher frequency resonance (blue) --- spatially shifted by $\lambda_B/2$ --- sensing the troughs, see inset of Fig. \ref{fig:OpticalSplitting}(b) and Supplemental Material.
The magnitude of the splitting provides the coupling rate $g_{\mathrm{opt}} =2 \pi \times 187$~MHz, which is larger than the half-width at half maximum of the optical mode ($\kappa/2  = 2\pi \times 142$~MHz), indicating mechanically mediated strong coupling between the two optical modes~\cite{Verhagen_Nat_2012}.

In summary, we have shown that optically induced surface deformations in liquids can facilitate greatly enhanced Brillouin interactions. We use this concept to demonstrate ultralow threshold Brillouin lasing in few-nanometer-thick superfluid helium films and, for the first time, phonon-mediated strong coupling between optical cavity modes.
While the microwatt-range lasing threshold presented here is already lower than all previously reported values
\cite{grudinin_brillouin_2009}, even modest improvements in acoustic dissipation and optical linewidth -- achieved for example by reducing sidewall roughness on our existing devices~\cite{Chemically} -- could lower the threshold power into the picowatt range, significantly below what can be achieved with solid-state systems.

The system presented here represents a substantial departure from traditional travelling-wave Brillouin systems, which typically exhibit weak single photon coupling due to large mode volumes and high Young's moduli. While this can be partially mitigated via resonant enhancement, doing so typically requires millimeter-scale devices to satisfy energy and momentum matching requirements~\cite{BrillouinGyroscope, MicrowaveSynthesizer}. In contrast, superfluid helium, with its ultra-compliant fluid interface, affords a low-frequency Brillouin shift that is naturally resonant with the co-propagating whispering gallery mode, regardless of the device size. This paves the way towards resonantly enhanced backward Brillouin scattering on devices with mode volumes as small as the optical wavelength cubed and with optomechanical coupling rates in excess of $g_{\mathrm{0,rp}}/2\pi\sim 300 $~kHz~\cite{baker_theoretical_2016}.

Phonon lasing in superfluid helium opens up a new approach to explore quantum turbulence and quantum fluid dynamics in a strongly-interacting system~\cite{sachkou_coherent_2019}. In this application, the combination of phonon-laser-induced mechanical line narrowing~\cite{feng_self-sustaining_2008} with operation as a precision Brillouin gyroscope~\cite{BrillouinGyroscope} would provide unprecedented sensitivity in measurements of the quantized circulation of the superfluid. Applied to conventional fluids, the same Brillouin process could be used to optically mix microdroplets and to probe their material properties, important tools for picolitre-scale chemistry, biophysics and the biosciences~\cite{destgeer_recent_2015, palombo_brillouin_2019}. Moreover, strong phonon-mediated optical coupling  may allow the generation of synthetic electromagnetic fields~\cite{Fang_NatPhys_17, Schmidt_Optica_15}, and superfluid based optical switches and reconfigurable optical circuits~\cite{Ruesink_NatComms_18, Shen_NatComms_18}.

\section*{Methods}

{\bf Experimental apparatus used to probe superfluid Brillouin lasing.} To experimentally probe superfluid thin-film Brillouin lasing, we employ the experimental apparatus shown in Fig. \ref{fig:setup}. The microdisk
is coupled to a tapered fiber and cooled to $\sim$20~mK in a sealed sample chamber within a dilution refrigerator. Helium-4 gas injected into the chamber forms a nanometer-thick self-assembling superfluid film coating the microresonator~\cite{harris_laser_2016} (see Supplemental Material). A telecom wavelength pump laser drives the Brillouin lasing process, and the backwards emitted Stokes and anti-Stokes light is observed using optical heterodyne detection and characterized on a spectrum analyser. We select a microdisk optical mode  at $\lambda_{\rm light}=1555$ nm with a loaded linewidth of $\kappa/2\pi = 284$~MHz, corresponding to an optical quality factor of  $Q\simeq 7\times 10^5$. This optical mode is also found to have a backscattering rate of $\kappa/2\pi=75$~MHz which suppresses the Brillouin induced linewidth narrowing by a factor of $\zeta=0.22$ (see Eq.~\ref{eq:Gammaeff2} and Supplemental Material). 

{\bf Pump-probe scheme used to characterize phonon-mediated strong coupling.} To spectroscopically characterize phonon-mediated strong coupling a weak tunable diode probe laser is added to the heterodyne setup shown in Fig.~\ref{fig:setup}. While the pump laser is set on resonance with the optical mode, the probe laser is repeatedly swept at 100~Hz across the optical resonance. The transmitted optical photodetector signal is low-pass filtered (filter bandwidth~=~20~kHz) and averaged 256 times on an oscilloscope in order obtain the optical cavity spectrum, free from any potential  modulation due to the Brillouin lasing process.

\section*{Acknowledgements}
This work was funded by the U.S. Army Research Office through grant number W911NF17-1-0310 and the Australian Research Council Centre of Excellence for Engineered Quantum Systems (EQUS, Project No. CE170100009). W.P.B. and  C.G.B respectively acknowledge Australian Research Council Fellowships FT140100650 and DE190100318. This work was performed in part at the Queensland node of the Australian National Fabrication Facility, a company established under the National Collaborative Research Infrastructure Strategy to provide nano and micro-fabrication facilities for Australia’s researchers.

\section*{References}

\footnotesize \renewcommand{\refname}{\vspace*{-30pt}}  
\bibliographystyle{ieeetr} 

\begin{thebibliography}{10}

\bibitem{Kittlaus_NatComm_17}
E.~A. Kittlaus, N.~T. Otterstrom, and P.~T. Rakich, ``On-chip inter-modal
  {B}rillouin scattering,'' {\em Nature Communications}, vol.~8, p.~15819,
  2017.

\bibitem{MicrowaveSynthesizer}
J.~Li, H.~Lee, and K.~J. Vahala, ``{Microwave synthesizer using an on-chip
  Brillouin oscillator},'' {\em Nature Communications}, vol.~4, no.~1, p.~2097,
  2013.

\bibitem{Eggleton_Review}
B.~J. Eggleton, C.~G. Poulton, and R.~Pant, ``Inducing and harnessing
  stimulated {B}rillouin scattering in photonic integrated circuits,'' {\em
  Adv. Opt. Photon.}, vol.~5, no.~4, pp.~536--587, 2013.

\bibitem{scarcelli2008confocal}
G.~Scarcelli and S.~H. Yun, ``Confocal brillouin microscopy for
  three-dimensional mechanical imaging,'' {\em Nature Photonics}, vol.~2,
  no.~1, p.~39, 2008.

\bibitem{Renninger_NatPhys_18}
W.~H. Renninger, P.~Kharel, R.~O. Behunin, and P.~T. Rakich, ``Bulk crystalline
  optomechanics,'' {\em Nature Physics}, vol.~14, no.~6, pp.~601--607, 2018.

\bibitem{HighFrequency}
P.~Kharel, G.~I. Harris, E.~A. Kittlaus, W.~H. Renninger, N.~T. Otterstrom,
  J.~G.~E. Harris, and P.~T. Rakich, ``High-frequency cavity optomechanics
  using bulk acoustic phonons,'' {\em Science Advances}, vol.~5, no.~4, 2019.

\bibitem{kashkanova2017superfluid}
A.~D. Kashkanova, A.~B. Shkarin, C.~D. Brown, N.~E. Flowers-Jacobs,
  L.~Childress, S.~W. Hoch, L.~Hohmann, K.~Ott, J.~Reichel, and J.~G.~E.
  Harris, ``Superfluid {B}rillouin optomechanics,'' {\em Nature Physics},
  vol.~13, no.~1, p.~74, 2017.

\bibitem{giorgini_stimulated_2018}
A.~Giorgini, S.~Avino, P.~Malara, P.~De~Natale, M.~Yannai, T.~Carmon, and
  G.~Gagliardi, ``Stimulated {Brillouin} {Cavity} {Optomechanics} in {Liquid}
  {Droplets},'' {\em Physical Review Letters}, vol.~120, no.~7, 2018.

\bibitem{Ruesink_NatComms_18}
F.~Ruesink, J.~P. Mathew, M.-A. Miri, A.~Alù, and E.~Verhagen, ``Optical
  circulation in a multimode optomechanical resonator,'' {\em Nature
  Communications}, vol.~9, no.~1, p.~1798, 2018.

\bibitem{Shen_NatComms_18}
Z.~Shen, Y.-L. Zhang, Y.~Chen, F.-W. Sun, X.-B. Zou, G.-C. Guo, C.-L. Zou, and
  C.-H. Dong, ``Reconfigurable optomechanical circulator and directional
  amplifier,'' {\em Nature Communications}, vol.~9, no.~1, p.~1797, 2018.

\bibitem{SafaviNaeini_NJP_11}
A.~H. Safavi-Naeini and O.~Painter, ``Proposal for an optomechanical traveling
  wave phonon{\textendash}photon translator,'' {\em New Journal of Physics},
  vol.~13, no.~1, p.~013017, 2011.

\bibitem{Fang_NatPhys_17}
K.~Fang, J.~Luo, A.~Metelmann, M.~H. Matheny, F.~Marquardt, A.~A. Clerk, and
  O.~Painter, ``Generalized non-reciprocity in an optomechanical circuit via
  synthetic magnetism and reservoir engineering,'' {\em Nature Physics},
  vol.~13, p.~465, 2017.

\bibitem{Schmidt_Optica_15}
M.~Schmidt, S.~Kessler, V.~Peano, O.~Painter, and F.~Marquardt,
  ``Optomechanical creation of magnetic fields for photons on a lattice,'' {\em
  Optica}, vol.~2, no.~7, pp.~635--641, 2015.

\bibitem{BrillouinGyroscope}
J.~Li, M.~Suh, and K.~Vahala, ``{Microresonator Brillouin gyroscope},'' {\em
  Optica}, vol.~4, no.~3, p.~346, 2017.

\bibitem{sachkou_coherent_2019}
Y.~P. Sachkou, C.~G. Baker, G.~I. Harris, O.~R. Stockdale, S.~Forstner, M.~T.
  Reeves, X.~He, D.~L. McAuslan, A.~S. Bradley, M.~J. Davis, and W.~P. Bowen,
  ``Coherent vortex dynamics in a strongly-interacting superfluid on a silicon
  chip,'' {\em arXiv:1902.04409 [cond-mat, physics:quant-ph]}, 2019.
\newblock arXiv: 1902.04409.

\bibitem{Gauthier1264}
G.~Gauthier, M.~T. Reeves, X.~Yu, A.~S. Bradley, M.~A. Baker, T.~A. Bell,
  H.~Rubinsztein-Dunlop, M.~J. Davis, and T.~W. Neely, ``Giant vortex clusters
  in a two-dimensional quantum fluid,'' {\em Science}, vol.~364, no.~6447,
  pp.~1264--1267, 2019.

\bibitem{ObservationOfBrillouin}
G.~Enzian, M.~Szczykulska, J.~Silver, L.~{Del Bino}, S.~Zhang, I.~A. Walmsley,
  P.~Del'Haye, and M.~R. Vanner, ``{Observation of Brillouin optomechanical
  strong coupling with an 11GHz mechanical mode},'' {\em Optica}, vol.~6,
  no.~1, p.~7, 2019.

\bibitem{Gundavarapu_IntegratedBrillouin}
S.~Gundavarapu, G.~M. Brodnik, M.~Puckett, T.~Huffman, D.~Bose, R.~Behunin,
  J.~F. Wu, T.~Q. Qiu, C.~Pinho, N.~Chauhan, J.~Nohava, P.~T. Rakich, K.~D.
  Nelson, M.~Salit, and D.~J. Blumenthal, ``Sub-hertz fundamental linewidth
  photonic integrated {B}rillouin laser,'' {\em Nature Photonics}, vol.~13,
  no.~1, p.~60, 2019.

\bibitem{Dong_BIT_reciprocity}
C.~H. Dong, Z.~Shen, C.~L. Zou, Y.~L. Zhang, W.~Fu, and G.~C. Guo,
  ``{B}rillouin-scattering-induced transparency and non-reciprocal light
  storage,'' {\em Nat Commun}, vol.~6, p.~6193, 2015.

\bibitem{Shin_NatCom_13}
H.~Shin, W.~Qiu, R.~Jarecki, J.~A. Cox, R.~H. Olsson~Iii, A.~Starbuck, Z.~Wang,
  and P.~T. Rakich, ``Tailorable stimulated {B}rillouin scattering in nanoscale
  silicon waveguides,'' {\em Nature Communications}, vol.~4, p.~1944, 2013.

\bibitem{VanLaer_NatPhot_2015}
R.~Van~Laer, B.~Kuyken, D.~Van~Thourhout, and R.~Baets, ``Interaction between
  light and highly confined hypersound in a silicon photonic nanowire,'' {\em
  Nature Photonics}, vol.~9, p.~199, 2015.

\bibitem{wolff_stimulated_2015}
C.~Wolff, M.~J. Steel, B.~J. Eggleton, and C.~G. Poulton, ``Stimulated
  {Brillouin} scattering in integrated photonic waveguides: {Forces},
  scattering mechanisms, and coupled-mode analysis,'' {\em Physical Review A},
  vol.~92, p.~013836, July 2015.

\bibitem{kaminski_ripplon_2016}
S.~Kaminski, L.~L. Martin, S.~Maayani, and T.~Carmon, ``Ripplon laser through
  stimulated emission mediated by water waves,'' {\em Nature Photonics},
  vol.~10, no.~12, pp.~758--761, 2016.

\bibitem{harris_laser_2016}
G.~I. Harris, D.~L. McAuslan, E.~Sheridan, Y.~Sachkou, C.~Baker, and W.~P.
  Bowen, ``Laser cooling and control of excitations in superfluid helium,''
  {\em Nature Physics}, vol.~12, no.~8, pp.~788--793, 2016.

\bibitem{baker_theoretical_2016}
C.~G. Baker, G.~I. Harris, D.~L. McAuslan, Y.~Sachkou, X.~He, and W.~P. Bowen,
  ``Theoretical framework for thin film superfluid optomechanics: towards the
  quantum regime,'' {\em New Journal of Physics}, vol.~18, no.~12, p.~123025,
  2016.

\bibitem{ObservationOfSpontaneous}
G.~Bahl, M.~Tomes, F.~Marquardt, and T.~Carmon, ``{Observation of spontaneous
  Brillouin cooling},'' {\em Nature Physics}, vol.~8, no.~3, pp.~203--207,
  2012.

\bibitem{Bahl_NatComm_2013}
G.~Bahl, K.~H. Kim, W.~Lee, J.~Liu, X.~D. Fan, and T.~Carmon, ``{B}rillouin
  cavity optomechanics with microfluidic devices,'' {\em Nature
  Communications}, vol.~4, 2013.

\bibitem{mcauslan_microphotonic_2016}
D.~L. McAuslan, G.~I. Harris, C.~Baker, Y.~Sachkou, X.~He, E.~Sheridan, and
  W.~P. Bowen, ``Microphotonic {Forces} from {Superfluid} {Flow},'' {\em
  Physical Review X}, vol.~6, no.~2, p.~021012, 2016.

\bibitem{ThirdSound}
K.~R. Atkins, ``{Third and fourth sound in liquid helium II},'' {\em Physical
  Review}, vol.~113, no.~4, pp.~962--965, 1959.

\bibitem{ashkin1986observation}
A.~Ashkin, J.~M. Dziedzic, J.~E. Bjorkholm, and S.~Chu, ``Observation of a
  single-beam gradient force optical trap for dielectric particles,'' {\em
  Optics letters}, vol.~11, no.~5, pp.~288--290, 1986.

\bibitem{Metzger_Nat_04}
C.~H. Metzger and K.~Karrai, ``Cavity cooling of a microlever,'' {\em Nature},
  vol.~432, no.~7020, pp.~1002--1005, 2004.

\bibitem{Jourdan_PRL_08}
G.~Jourdan, F.~Comin, and J.~Chevrier, ``Mechanical mode dependence of
  bolometric backaction in an atomic force microscopy microlever,'' {\em
  Physical Review Letters}, vol.~101, no.~13, 2008.

\bibitem{Verhagen_Nat_2012}
E.~Verhagen, S.~Deleglise, S.~Weis, A.~Schliesser, and T.~J. Kippenberg,
  ``Quantum-coherent coupling of a mechanical oscillator to an optical cavity
  mode,'' {\em Nature}, vol.~482, no.~7383, pp.~63--67, 2012.

\bibitem{grudinin_brillouin_2009}
I.~S. Grudinin, A.~B. Matsko, and L.~Maleki, ``{B}rillouin {Lasing} with a
  {CaF}$_2$ {Whispering} {Gallery} {Mode} {Resonator},'' {\em Physical Review
  Letters}, vol.~102, no.~4, p.~043902, 2009.

\bibitem{Chemically}
H.~Lee, T.~Chen, J.~Li, K.~Y. Yang, S.~Jeon, O.~Painter, and K.~J. Vahala,
  ``{Chemically etched ultrahigh-Q wedge-resonator on a silicon chip},'' {\em
  Nature Photonics}, vol.~6, no.~6, pp.~369--373, 2012.

\bibitem{feng_self-sustaining_2008}
X.~L. Feng, C.~J. White, A.~Hajimiri, and M.~L. Roukes, ``A self-sustaining
  ultrahigh-frequency nanoelectromechanical oscillator,'' {\em Nature
  Nanotechnology}, vol.~3, pp.~342--346, June 2008.

\bibitem{destgeer_recent_2015}
G.~Destgeer and H.~J. Sung, ``Recent advances in microfluidic actuation and
  micro-object manipulation via surface acoustic waves,'' {\em Lab on a Chip},
  vol.~15, pp.~2722--2738, June 2015.

\bibitem{palombo_brillouin_2019}
F.~Palombo and D.~Fioretto, ``Brillouin {Light} {Scattering}: {Applications} in
  {Biomedical} {Sciences},'' {\em Chemical Reviews}, vol.~119, pp.~7833--7847,
  July 2019.

\end{thebibliography}

\end{document}


\beginsupplement

\title{Strong optical coupling through superfluid Brillouin lasing - Supplementary Information}

\author{Xin He, Glen I. Harris, Christopher G. Baker, Andreas Sawadsky,\\ Yasmine L. Sfendla, Yauhen P. Sachkou, Stefan Forstner, Warwick P. Bowen}

\date{}


\maketitle

\tableofcontents

\section{Experimental details}
\label{section_experimental_details}
\subsection{Device Fabrication}

Silica microdisks are fabricated from a 500 $\mu$m-thick silicon handling wafer topped by a two-micron thick thermal oxide layer (Virginia Semiconductor). Disks are defined in the silica layer through a combination of photolithography (AZ1518 positive resist and HMDS adhesion promoter) and hydrofluoric acid (HF) wet-etch. A subsequent XeF$_2$ gas-phase release selectively etches the silicon material and leaves the silica disks isolated from the substrate atop a silicon pedestal, as shown in Fig. 1 of the main text.

Unlike in previous experimental work~\cite{harris_laser_2016, mcauslan_microphotonic_2016}, the fabricated microdisks undergo no laser reflow step to form a microtoroidal resonator, and maintain their wedged outer sidewalls. This wedge shape serves a dual purpose: beyond enhancing the optical Q (over a vertical sidewalled microdisk) by making the device less sensitive to fabrication-induced roughness~\cite{lee_chemically_2012}, it also serves to deconfine the optical mode and maximize the optical  field intensity at the top and bottom disk interfaces where the superfluid film resides, as shown in Fig. \ref{Figfilmthickness}. Indeed, the value of the WGM electric field at the silica interface is the parameter which should be optimized in order to maximize the optomechanical coupling rate between light and superfluid~\cite{baker_theoretical_2016}, see section \ref{sectioncomputationofoptomechanicalcouplingrateG}. Fabricated devices show a number of WGM families, with optical Qs in the $10^5$ to low $10^7$ range.

\subsection{Experimental setup}

The microresonator chip is positioned inside a superfluid-tight sample chamber at the bottom of a
Bluefors dilution refrigerator (base temperature 10 mK). 
Laser light is evanescently coupled into the microresonators via a tapered optical fiber \cite{harris_laser_2016}. Precise fiber positioning is achieved through Attocube nanopositioning stages. The fiber taper rests on support pads microfabricated on the chip alongside the resonators~\cite{mcauslan_microphotonic_2016}, in order to eliminate taper drift and fluctuations. The experimental measurements are performed with the pulse-tube cooler switched off to minimize vibrations. The sample chamber contains a small volume of alumina nanopowder in order to increase the  effective chamber surface area (by $\sim 10$ m$^2$), leading to more precise film thickness control and greater film thickness stability \cite{ellis_observation_1989}.
While at base temperature, $^4$He gas can be continuously injected from the top of the cryostat into the sample chamber through a thin capillary. This allows for varying of the film thickness, and thus in-situ tuning of the Brillouin frequency.

\subsection{Characterization of superfluid film thickness}

The thickness of the superfluid film covering the microresonator can be assessed through two independent means:
\begin{itemize}
\item First, through the magnitude of the WGM wavelength-shift across the superfluid transition temperature. When sweeping the cryostat temperature down from 1.1 K to 0.2 K, all WGMs acquire a positive wavelength shift corresponding to the increased optical path length due to the condensation  of the superfluid film on the resonator surface\footnote{Indeed, the magnitude of the thermo-optic shift~\cite{gil-santos_light-mediated_2017} over this temperature range can be neglected.}, as shown in Fig. \ref{Figfilmthickness}. 
\begin{figure}
\centering
\includegraphics[width=\textwidth]{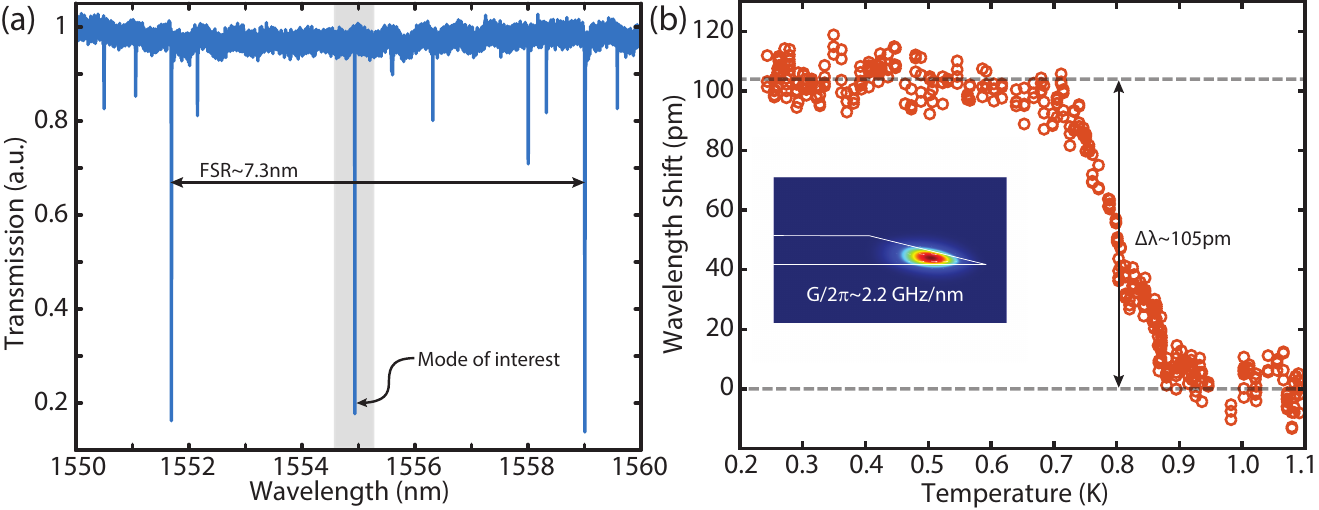}
\caption{a)  Microdisk optical spectrum, showing a number of high-Q WGMs separated by a $\sim7$ nm free-spectral-range (FSR). WGM highlighted in gray near 1555 nm is the one used in the experiments.   b) WGM wavelength shift as a function of fridge temperature, for the mode highlighted in (a). The vertical scatter in the data points is due to the 20 pm repeatability error in the motor sweep of our tunable laser diode. Inset displays the electric field norm of the $(p=1,\,m_{\mathrm{opt}}=186)$ quasi-TE mode of the structure. } 
\label{Figfilmthickness}
\end{figure}
Each WGM experiences a frequency shift $\frac{G}{2\pi} \Delta x$, where $G=\frac{\partial \omega_0}{\partial x}$ is the optomechanical coupling strength describing the optical cavity angular resonance frequency shift per unit deposited superfluid film thickness on the resonator boundary $\Delta x$~\cite{baker_theoretical_2016, Aspelmeyer2014}. 
The value of G is WGM dependent, and can be calculated through FEM modelling, as detailed in section \ref{sectioncomputationofoptomechanicalcouplingrateG} below. With the knowledge of G, the magnitude of the experimentally measured WGM frequency shift can be converted into a deposited film thickness.

\item Second, through the Brillouin frequency. The Brillouin wavelength is imposed by the wavelength of light through $\lambda_B\approx\lambda_{\mathrm{light}}/2$,  where $\lambda_{\mathrm{light}}$ is the wavelength of light in the silica given by $\lambda_0/n_{\mathrm{eff}}$, \textit{i.e.} the freespace wavelength $\lambda_0$ divided by the WGM effective index. The Brillouin frequency $\ob/2\pi=c_3/\lambda_B$ thus informs us on the speed of sound in the superfluid $c_3$, which is given by~\cite{atkins_third_1959}:
\begin{equation}
c_3=\sqrt{3\, \frac{\rho_s}{\rho}\frac{\alpha_{\mathrm{vdw}}}{d^3}}.
\label{Eqspeedofsoundc3}
\end{equation}
Here the ratio $\rho_s/\rho$ is nearly one at the low temperature used in our experiments, $\alpha_{\mathrm{vdw}}=2.6\times10^{-24}$ $\mathrm{m}^5\, \mathrm{s}^{-2}$ describes the van der Waals interaction between the helium atoms in the superfluid and the silica disk~\cite{baker_theoretical_2016} and $d$ is the superfluid film thickness. The experimentally measured Brillouin frequency thus provides a second independent estimate of the film thickness in the experiments.
\end{itemize}
The film thickness estimation through these two methods is provided in Table \ref{tableexperimentalparameters}. We ascribe the discrepancy to uncertainties in the exact device geometry. Indeed, variations in wedge angle of a few degrees can shift $G$ by over 30\%, by altering the mode confinement of the WGM and its interaction with the superfluid film. Surface roughness on the microresonator, not taken into account in the simulations, may also increase the effective surface area of the resonator and result in uncertainties in the $G$.

\subsection{Optomechanical coupling $G$}
\label{sectioncomputationofoptomechanicalcouplingrateG}

We compute the optomechanical coupling strength $G=\frac{\partial \omega_0}{\partial x}$ using FEM modelling software (COMSOL Multiphysics).
The silica microresonator dimensions are measured with a scanning electron microscope (SEM) and summarized in Table \ref{tableexperimentalparameters}. We simulate the optical eigenmodes of the structure, which for thin disks are defined by their transverse electric (TE) or transverse magnetic (TM) polarization, and radial and azimuthal mode orders $(p,m)$ \cite{ding_high_2010}. The electric field ($E$) distribution of the $(p=1,\, m=186)$ quasi-TE mode of the microdisk is plotted as an inset in Fig. \ref{Figfilmthickness}(b). $G$ is computed from the $E$ field distribution through~\cite{baker_theoretical_2016}:
\begin{equation}
G=\frac{-\omega_0}{2}\,\frac{\iint_{\mathrm{interface}} \left( \varepsilon_{\mathrm{sf}} -1 \right)E^2\left( \vec{r} \right) \mathrm{d}^2\vec{r}}{\iiint_{\mathrm{all}} \varepsilon_r\left( \vec{r}\right) E^2\left( \vec{r} \right) \mathrm{d}^3\vec{r}},
\label{EquationG}
\end{equation}
where $\varepsilon_r$ is the relative permittivity and $\varepsilon_{\mathrm{sf}}=1.058$ is the relative permittivity of superfluid helium. The numerator surface integral is performed over both top and bottom resonator boundaries, while the normalizing denominator volume integral is performed over all space. We find values of $G/2\pi$ clustered around $-2.2 \times 10^{18}$ Hz/m for TE modes and around $-2.4 \times 10^{18}$ Hz/m for TM modes, with little influence of the WGM radial order. Higher values of $\abs{G}$ for TM modes is due to their stronger field at the interface due to the $E$ field discontinuity \cite{baker_theoretical_2016}. We identify our experimental WGM as a $(p=1,m_{\mathrm{opt}}=186)$ quasi-TE WGM, as shown in the inset of Fig. \ref{Figfilmthickness}(b).
As mentioned above, the optomechanical coupling $G$ has a marked dependence on the wedge angle. Indeed, while the fundamental TE mode's $\abs{G}/2\pi\sim 2.2$ GHz/nm for a 14 degree wedge angle, this value increases to 2.5 GHz/nm for a 12 degree wedge, and drops to 1.5 GHz/nm for a 20 degree wedge angle.
The contribution to the total G from top and bottom disk interfaces is well balanced, with respectively 51\% and 49 \% of the total coupling rate coming from top and  bottom  for the fundamental TE mode. 

\begin{table}
\centering
    \begin{tabular}{l|c|c|c|c}
        \hline
        Parameter                & Symbol       & Value  & Units   & source      \\
           \hline
           \hline
               Disk radius (top)               & $R_t$                & $ 30.6 $  & $\mu$m  & SEM \\ 
    Disk radius (bottom)               & $R_b$                & $ 38.6 $  & $\mu$m  & SEM \\ 
     Disk wedge angle                &       -          & $14 $  &  degrees  & SEM \\ 
    Film thickness              & $d $                & $6$    &  nm & cavity shift \\ 
                 &                & $8$    &  nm & Brillouin frequency \\ 
                 Speed of sound & $c_3$ & $\sim 5$ & m/s & Eq.(\ref{Eqspeedofsoundc3})\\
    WGM azimuthal number        & $m_{\mathrm{opt}} $                & $186 $ & - & FEM \\ 
    Mechanical azimuthal number        & $m $                & $372 $ & -  & -\\ 
    Optomechanical coupling rate  & $G/2\pi$ & $-2.17 \times 10^{18}$  & Hz/m & FEM \\ 
    Brillouin mode zero-point motion                   & $x_{\mathrm{ZPF}} $   & $9.5 \times 10^{-15}$  & m & analytical estimation  \\
        Single photon coupling strength             & $g_{\mathrm{0,tot}}/ 2 \pi $                    & $133$   & $ \mathrm{kHz}$ & Numerical fit to measurement\\
                  &       $g_{\mathrm{0,rp}}/ 2 \pi $            & $11$   & $ \mathrm{kHz}$ & simulation\\
									&       $g_{\mathrm{0,fp}}/ 2 \pi $            & $122$   & $ \mathrm{kHz}$ & Numerical fit to measurement\\
        \hline
    \end{tabular}
\caption{Experimental parameters. SEM: Scanning Electron Microscope; FEM: Finite Element Method.}
\label{tableexperimentalparameters}
\end{table}

\subsection{Estimation of single photon optomechanical coupling rate $g_{0,\mathrm{rp}}$}
\label{sectionestimationofg0}

\begin{figure}
\centering
\includegraphics[width=.75\textwidth]{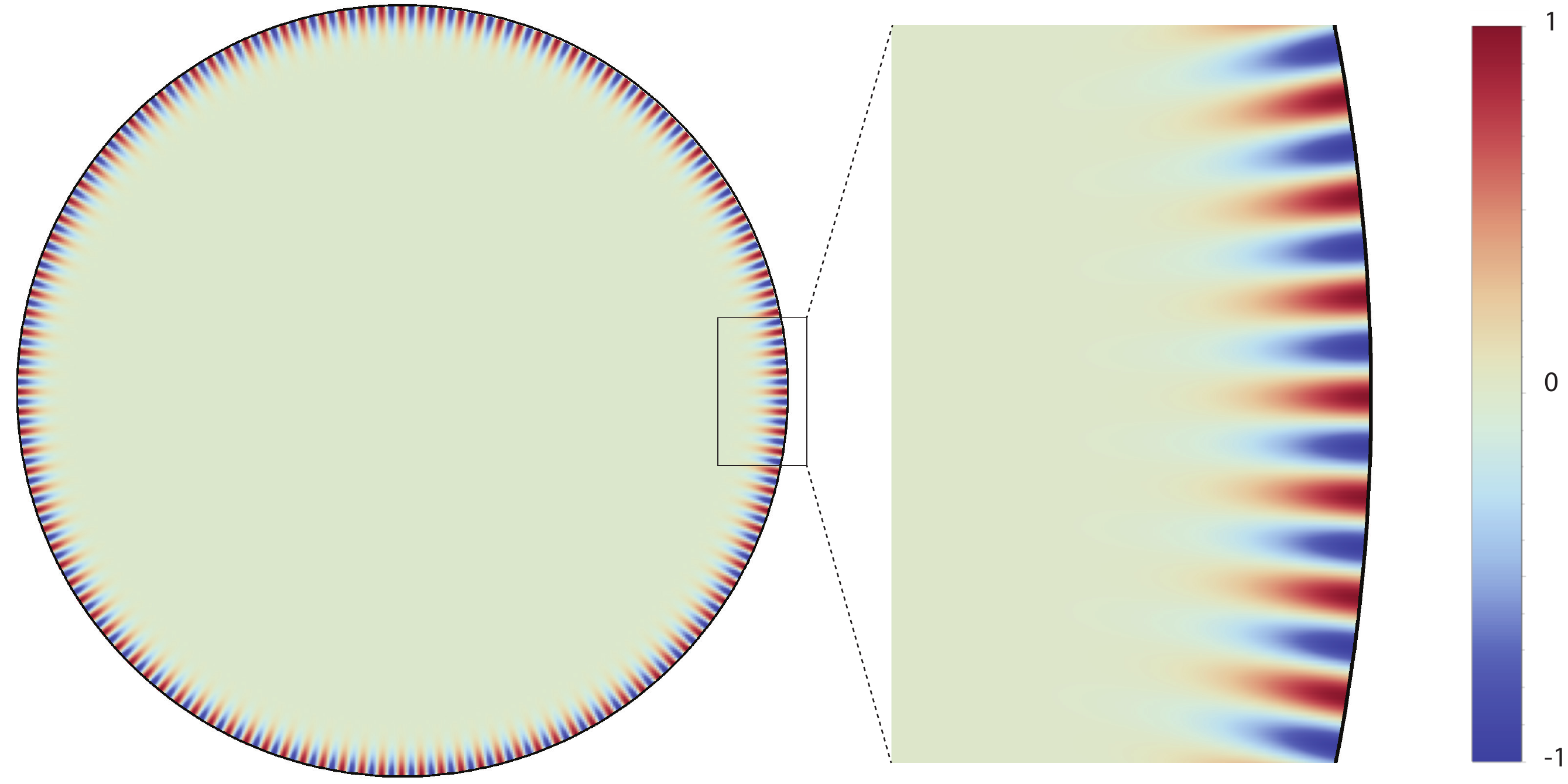}
\caption{
Example out-of-plane displacement profile of an acoustic whispering gallery-type mode (superfluid Brillouin wave) of the kind used in the experiments. For clarity, we show here the normalized displacement $\eta_{120,1}\left( r, \theta\right)$ of the $m=120$ Bessel mode with free boundary conditions at the resonator edge \cite{sachkou_coherent_2019}, which has $\sim3$ times fewer nodes than the experimental $m=372$ Bessel mode. The displacement is essentially localized on the resonator edge, such that the eigenmode is not perturbed by the presence of the pedestal on the microdisk underside.
 } 
\label{Figmechmode}
\end{figure}

Calculation of the single-photon optomechanical coupling rate $g_{0,\mathrm{rp}}=\abs{G} \, x_{\mathrm{ZPF}}$~\cite{Aspelmeyer2014} requires the Brillouin mode zero-point motion $x_{\mathrm{ZPF}}$.
The Brillouin eigenmode can be well approximated through a high-order azimuthal Bessel mode of a disk, for which the zero-point motion can be calculated analytically \cite{baker_theoretical_2016}, see Fig. \ref{Figmechmode}. 
The surface displacement $\eta$ of the travelling Brillouin wave on both top and bottom surfaces of the disk is given by:
\begin{equation}
\eta_{m,n}\left( r,\theta, t \right)=\eta_0\, J_m\left( \zeta_{m,n} \frac{r}{R} \right) \cos\left( m \theta  \pm \ob\,t\right), 
\end{equation}
where $m$ and $n$ are respectively the azimuthal and radial mode numbers, $\eta_0$ the mode amplitude, $J_m$ the Bessel
function of the first kind of order $m$, and $\zeta_{m,n}$ a frequency parameter depending on the mode order and the boundary conditions \cite{baker_theoretical_2016}.
Energy and momentum conservation for the Brillouin scattering process imply that the Brillouin mode azimuthal order be twice that of the optical WGM, \textit{i.e.} $m=2\,m_{\mathrm{opt}}=2\times 186=372$ (see Table \ref{tableexperimentalparameters}). Such a mode has its displacement localized on the periphery of the resonator, well colocalized with the optical field intensity,  forming a type of acoustic whispering gallery mode, as shown in Fig. \ref{Figmechmode}. Note that since the excitations exist on both the top and underside of the disk, the collective excitation of the film on top and bottom has twice the effective mass, and hence $1/\sqrt{2}$ the zero-point motion of a mode residing only on the disk top surface. The azimuthal overlap between optical and mechanical fields leads to a further factor two reduction: $\frac{1}{2\pi}\int_0^{2\pi} 2 \cos^2\left( m_{\mathrm{opt}} \right) \cos\left(m\right)=\frac{1}{2}$. 
Combined, this estimation provides, for the WGM mode used in the experiments a value of $g_{\mathrm{0,rp}}= 11$ kHz for the radiation pressure component of the single photon coupling rate. 
The fountain pressure contribution to the optical forcing, estimated to be $g_{\mathrm{0,fp}}= 122$ kHz from fitting to numerical simulations, arises from superfluid flow induced from optical heating~\cite{mcauslan_microphotonic_2016}. As discussed in the main text, feasible improvements to the device (i.e. reduced optical and mechanical dissipation and increased optomechanical coupling through changes in disk material and thickness) can both  suppress fountain pressure contributions and substantially increase radiation pressure effects, allowing the latter to dominate.

\subsection{Influence of surface tension}

\begin{figure}
\centering
\includegraphics[width=0.5\textwidth]{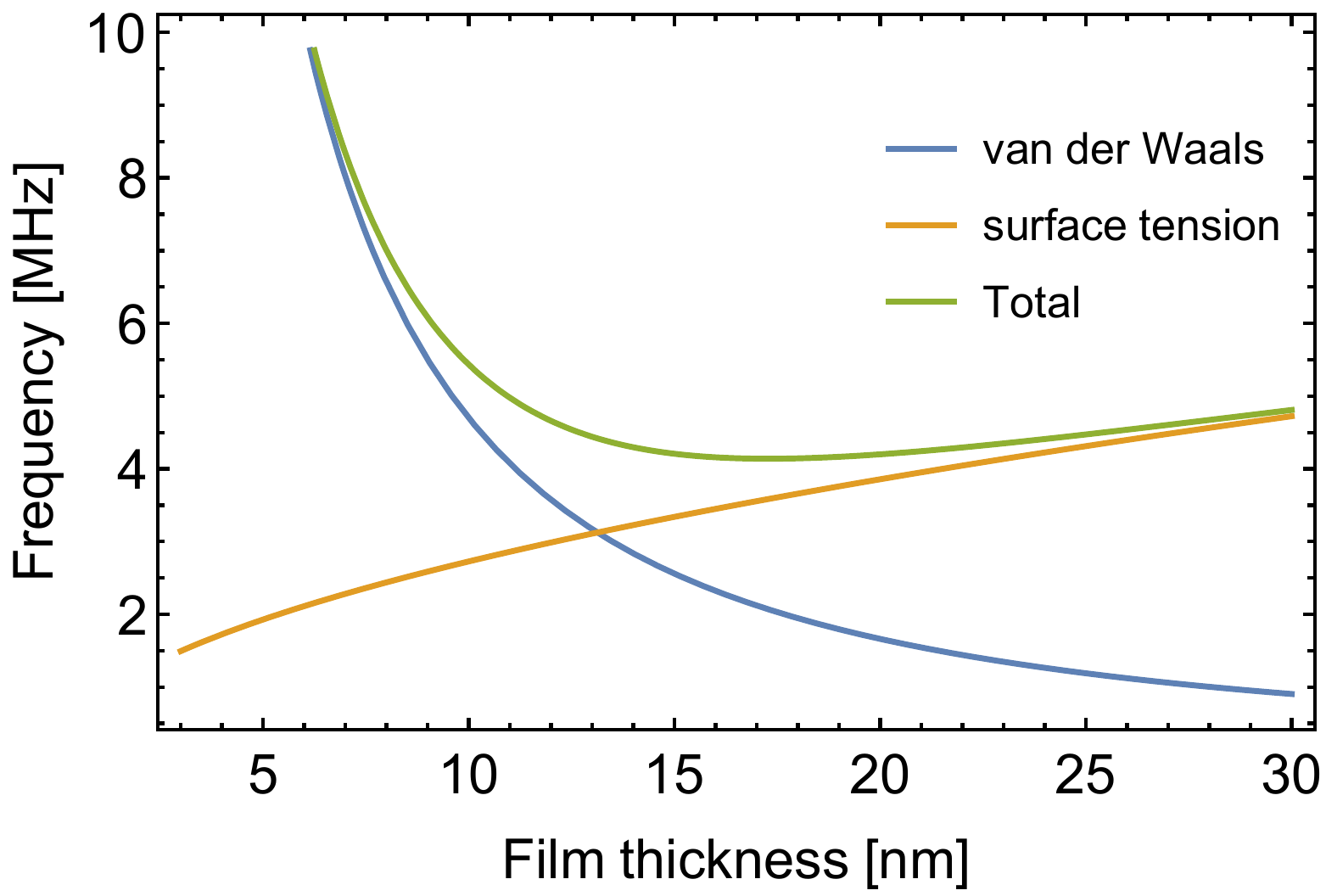}
\caption{Frequency $\ob/2\pi$ of the Brillouin wave in the presence of both van der Waals and surface tension restoring forces (green), as well as the limiting cases of the frequency $\Omega_{\mathrm{vdw}}/2\pi$ of a pure van der Waals wave (third sound - blue) and the frequency $\Omega_{\sigma}/2\pi$ of a  pure surface tension wave (ripplon - orange).  Values plotted with $k=10^7$, corresponding to the Brillouin wavenumber in our experiments.} 
\label{Figsurfacetension}
\end{figure}
The dispersion relation giving the angular frequency $\Omega$ of a superfluid wave under the influence of van der Waals and surface tension restoring forces is given  by:
\begin{equation}
\Omega=\sqrt{\frac{3\,\alpha_{\mathrm{vdw}}\,k^2}{d^3}+\frac{\sigma\,k^4\,d}{\rho}},
\end{equation}
where $\sigma=3.54 \times 10^{-4}$ N/m is the superfluid $^4$He surface tension~\cite{donnelly_observed_1998} and $k=2\pi/\lambda=\zeta_{m,n}/R$ the angular wavenumber. 
We plot this frequency $\Omega/2\pi$ in Fig. \ref{Figsurfacetension}, along with the limiting cases of a pure van der Waals wave (third sound) of frequency $\Omega_{\mathrm{vdw}}/2\pi=\frac{1}{2\pi}\,\sqrt{\frac{3\,\alpha_{\mathrm{vdw}}\,k^2}{d^3}}$, and a pure surface tension wave (ripplon) of frequency  $\Omega_{\sigma}/2\pi=\frac{1}{2\pi}\,\sqrt{\frac{\sigma\,k^4\,d}{\rho}}$. For the film thickness used in the experiment, the dominant restoring force is the van der Waals interaction, and the Brillouin wave can be well approximated by a third sound mode. 
For thicker films ($>13$ nm), the restoring force becomes dominated by surface tension, and the wave crosses over into a ripplon-like regime.

\section[Standing-wave versus travelling-wave Brillouin interaction]{Standing-wave versus travelling-wave Brillouin interaction}
\label{sectionBrillouinvsoptomechanics}

\begin{figure}
\centering
\includegraphics[width= 0.65\textwidth]{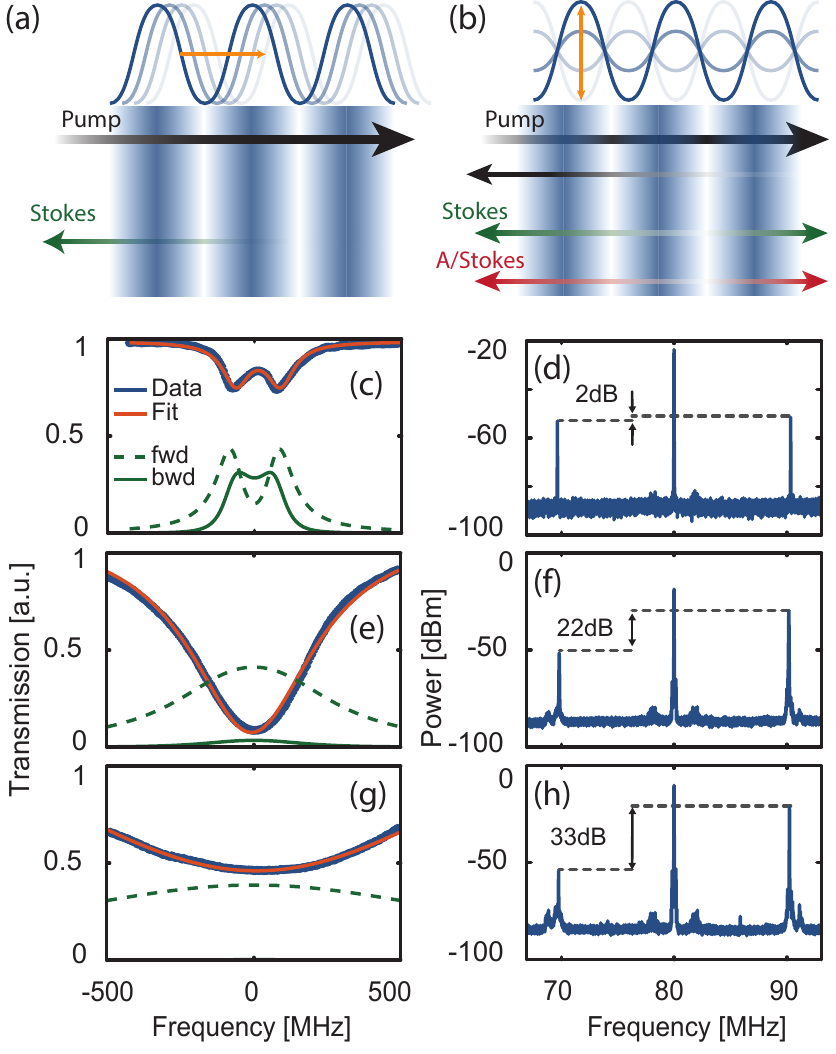}%
\caption{(a) Schematic illustration of a travelling wave `Brillouin-like' interaction, where the pump scatters off a moving refractive index grating, leading to single Stokes sideband generation. (b) `Optomechanics-like' interaction where the pump is scattered by a standing refractive index grating whose strength is modulated in time, leading to symmetric Stokes and anti-Stokes sideband generation. (c),(e),(g): normalized experimental WGM transmission spectra for increasing fiber taper coupling strength (blue line). The dashed and solid green lines respectively refer to the photon numbers of the forward and backward travelling directions $\abs{a_k}^2$ and $\abs{a_{-k}}^2$ (arbitrary units). By increasing the coupling the optical field changes from predominantly standing to predominantly travelling:  
$\abs{a_k}^2/\abs{a_{-k}}^2$ at zero detuning goes from 0.53 (c) to 11.7 (e) and 162.4 (g). (d), (f) and (h): experimental heterodyne power spectra respectively corresponding to the cases (c), (e) and (g), illustrating the transition from symmetric to asymmetric sideband generation.}%
\label{fig:sidebands}%
\end{figure}

There has been a recent push to theoretically unify the fields of cavity optomechanics and Brillouin scattering~\cite{Unifying}. While both fields deal with the inelastic interaction of photons with a mechanical degree of freedom (phonons), the Brillouin scattering paradigm often refers to the interaction between a travelling optical field and a travelling mechanical wave which causes a periodic refractive index modulation (see Fig. \ref{fig:sidebands}(a)). In contrast,  the optomechanical paradigm --as illustrated by the archetypal Fabry-Perot cavity with a movable end-mirror-- typically refers to the interaction between a standing optical field and a standing mechanical wave, as illustrated in Fig. \ref{fig:sidebands}(b).

We show here that these two regimes can be accessed on the same device, and that the switch can be performed in-situ, simply by tuning the position of the coupling fiber taper in order to transition from a standing to a travelling intracavity optical field. Indeed, our microresonator possesses some native optical backscattering due to geometric imperfections such as sidewall roughness, which introduces a coupling between forward and backward propagating directions.
This rate is experimentally measured to be $\kappa_b=75$ MHz, on the order of the intrinsic linewidth $\kappa_{int}=104$ MHz, and manifests as the optical resonance taking on a characteristic doublet lineshape
 \cite{Borselli2005}, as shown in the blue trace in Fig. \ref{fig:sidebands}(c).
 
Based on coupled-mode theory formalism (see section \ref{sectionCMTdoublet} below), one can compute the relative amount of light travelling in the forward and backward directions as a function of detuning in this regime, as shown in the green traces.
Due to the backscattering rate $\kappa_b$ being of comparable magnitude to the loss rate $\kappa$, both circulation directions are similarly populated, leading to a predominantly standing optical field, as described in Fig. \ref{fig:sidebands}(b).
Indeed, we verify that in this regime the Stokes and anti-Stokes sidebands are comparable in magnitude, as shown in Fig. \ref{fig:sidebands}(d).

Next, we increase the coupling rate $\kappa_{ext}$ of the cavity by approaching the fiber taper. As the cavity was previously undercoupled, this increases the depth of the transmission dip, while broadening the width of the resonance, as shown in Fig. \ref{fig:sidebands}(e).
In this regime, the backscattering rate $\kappa_b$ is no longer larger than the linewidth $\kappa=\kappa_{int}+\kappa_{ext}$, and the forward propagating field is predominantly populated (green curves), leading to a predominantly travelling optical field.
This can be understood by looking at Eqs. (\ref{EqakCMTdoublet}) and (\ref{EqaminuskCMTdoublet}) in the following, which describe the intracavity field amplitudes. Both propagation directions experience a loss rate $\kappa=\kappa_{int}+\kappa_{ext}$ which depends on $\kappa_{ext}$. However, while increasing $\kappa_{ext}$ increases both the loss rate and the pump rate of the forward field (see Eq. (\ref{EqakCMTdoublet})), it only increases the loss rate of the backward propagating mode, which is pumped at a fixed rate proportional to $\kappa_b$ (see Eq. (\ref{EqaminuskCMTdoublet})). Increasing the coupling rate to the taper therefore biases the system towards a forward travelling optical field and a situation analogous to the Brillouin case described in Fig. \ref{fig:sidebands}(a). Indeed, in this coupling regime the asymmetry between Stokes and anti-Stokes sidebands reaches 22 dB, as shown in Fig. \ref{fig:sidebands}(f).
Further increasing the taper coupling rate well into the overcoupled regime (Fig. \ref{fig:sidebands}(g)) leads to an even higher ratio of forward to backward optical intensity ($\abs{a_k}^2/\abs{a_{-k}}^2=162$ on resonance), and a sideband asymmetry reaching 33 dB (Fig. \ref{fig:sidebands}(h)).

\subsection{Hamiltonian of optical fields in the presence of backscattering}

In addition to a mechanically mediated coupling (e.g. the Brillouin induced strong coupling discussed in section \ref{sectionopticalstrongcoupling}), the coupling of two WGMs can be induced by the backscattering caused by imperfections of the resonator geometry such as sidewall roughness \cite{Borselli2005}. Unlike the mechanically induced optical coupling, this type of coupling is stationary, so we define a fixed mutual coupling strength $\kappa_b$ between the two optical modes $a_k$ and $a_{-k}$ (forward and backward propagating respectively). The Hamiltonian of the system derived from coupled mode theory is:
\begin{equation}
\label{eq:Hio}
\hat H  = \hbar \omega_k a^\dagger_k a_k + \hbar \omega_{-k}a^\dagger_{-k}a_{-k} - \hbar \kappa_b \left (a^\dagger_k a_{-k} + a_k a^\dagger_{-k}  \right ),
\end{equation}
where $\omega_k$ and $\omega_{-k}$ are the bare eigen-frequencies of the two optical modes, and the associated $a$ is the lowering operator of each mode.  The subscript $k$ is the wave number of the optical modes, with the signs in the front indicating direction of propagation. The interaction term is a beamsplitter interaction between the two optical modes. In the case of an ideal WGM resonator $\omega_k = \omega_{-k}$, but in the presence of strong backscattering (i.e. $\kappa_b>>\kappa$) the bare eigen-modes of the cavity hybridize into two distinct non-degenerate modes~\cite{Borselli2005}.

\subsection{Equations of motion}
\label{sectionCMTdoublet}
Including dissipation and drive, but ignoring vacuum fluctuations, the equations of motion are obtained for the coupled optical cavity modes from the Hamiltonian of Eq.(\ref{eq:Hio}): 
\begin{eqnarray}
\dot a_k &=&  i \Delta a_k -\frac{\kappa}{2}a_k + i \kappa_b a_{-k} +\sqrt{\kappa_{ext}}a_{in} \label{EqakCMTdoublet}\\
\dot a_{-k} &=& i \Delta a_{-k} -\frac{\kappa}{2}a_{-k} + i \kappa_b a_k,  \label{EqaminuskCMTdoublet}
\end{eqnarray}
where $\Delta = \omega_L - \omega_{k\left(-k\right)}$ is the detuning, $\kappa = \kappa_{int} + \kappa_{ext}$ is the total optical decay rate for either optical mode. $\kappa_{ext}$ is the coupling rate to the tapered optical fibre, and $\kappa_{int}$ is the intrinsic  optical cavity decay rate. The pump field $a_{in}$ drives the forward propagating mode $a_k$ via the fiber taper. The amplitude of the pump field $a_{in}$ is related to the incoming photon flux in the tapered fibre, via $\abs{a_{in}}^2 = \frac{P_{in}}{\hbar \omega}$, with $P_{in}$ the input laser power. Solving the equations of motion in the steady state, the forward propagating mode $a_k$ has the solution:
\begin{equation}
\label{eq:akbackscattering}
a_k = \frac{\sqrt{\kappa_{ext}}\,a_{in}}{-i\Delta + \kappa_k/2 + \frac{\kappa_b^2}{-i\Delta + \kappa_{-k}/2}}.
\end{equation}
The amplitude of the backscattered field $a_{-k}$ travelling in the opposite direction can be expressed as a function of $a_k$:
\begin{equation}
\label{eq:aminuskbackscattering}
a_{-k} = \frac{i \, \kappa_b \, a_k}{-i \Delta + \kappa_{-k}/2}.
\end{equation}
Using the input-output theorem, the output light in the fibre after the cavity is $a_{out}$, equal to $a_{in} - \sqrt{\kappa_{ext}}\,a_k$. Thus, the normalised transmission spectrum in the fibre after the WGM cavity is:
\begin{equation}
\label{eq:T}
\begin{split}
T &= \left|\frac{a_{out} }{a_{in}}\right|^2\\
 &=\left|1-\frac{\sqrt{\kappa_{ext}}a_k}{a_{in}}\right|^2
\end{split}
\end{equation}

\begin{figure}
\centering
\includegraphics[width=0.95\textwidth]{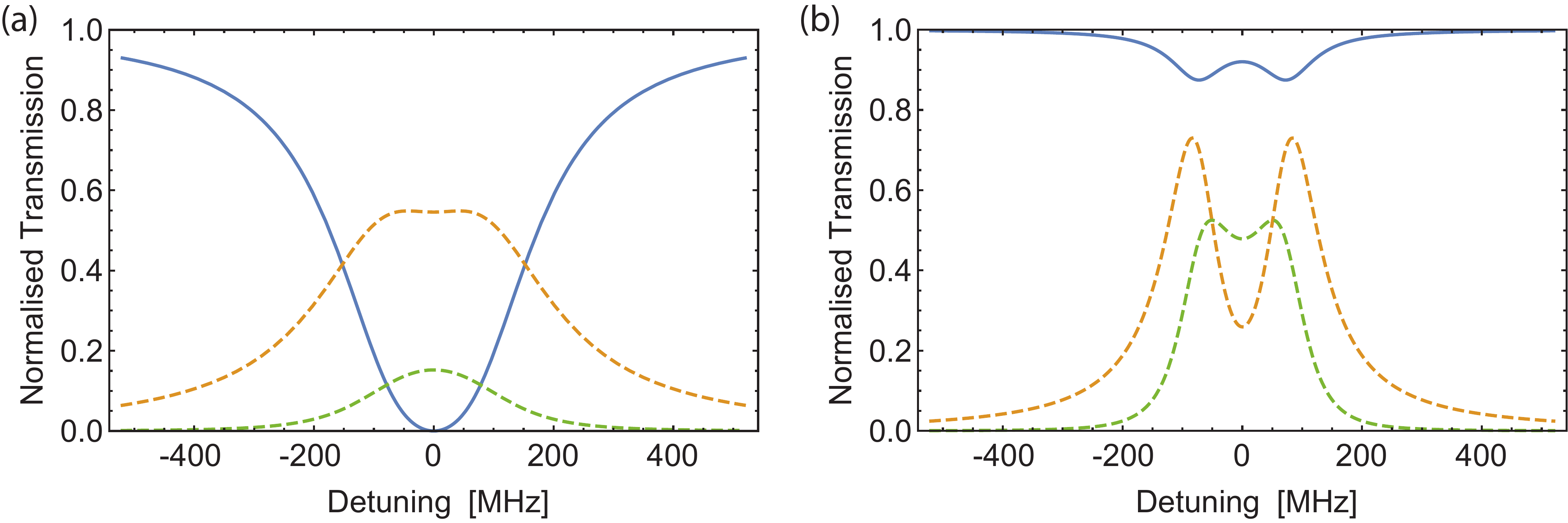}%
\caption{Normalised WGM transmission spectrum in the presence of backscattering, plotted using Eq.~(\ref{eq:T}). Blue: Normalized transmission. Yellow and green dashed lines refer respectively to the forward and backward propagating field intensities $|a_k|^2$ and $|a_{-k}|^2$ in arbitrary units, obtained from Eqs (\ref{eq:akbackscattering}) and (\ref{eq:aminuskbackscattering}). (a) Intrinsic optical linewidth $\kappa_{int}$: 104 MHz, taper coupling rate $\kappa_{ext}$: 180 MHz, backscattering rate $\kappa_b$: 75 MHz. (b) Undercoupled regime with intrinsic optical linewidth $\kappa_{int}$: 104 MHz, taper coupling rate $\kappa_{ext}$: 6.4 MHz, backscattering rate $\kappa_b$: 75 MHz. }%
\label{fig:T}%
\end{figure}

The effect of the taper coupling rate $\kappa_{ext}$ on the ratio of forward to backward travelling light intensity is illustrated in Fig.~\ref{fig:T}.
We consider an optical mode with an unloaded optical decay rate $\kappa_{int}=104$ MHz, and a backscattering coupling rate $\kappa_b=75$ MHz. When the taper coupling rate is set to $\kappa_{ext}=180$ MHz, the resulting  cavity transmission is plotted in Fig.~\ref{fig:T}(a). Because the total dissipation rate $\kappa=\kappa_{int}+\kappa_{ext}$ is larger than $\kappa_b$, the mode splitting is not resolvable in the cavity transmission (blue curve). For the same reason, the intracavity intensity of the forward propagating optical field (dashed yellow line) is around 4 times larger than that of the field propagating in the opposite direction (dashed green line). This intensity difference of the two optical modes leads to a net forward travelling optical field in the cavity. In contrast, when the taper coupling rate is reduced such that the total dissipation rate is now comparable to the backscattering rate, the lifted degeneracy between forward and backward propagating fields is revealed, as shown in the blue transmission trace in Fig.~\ref{fig:T}(b). In this regime,  both optical fields are similarly populated, resulting in a predominantly standing optical field.

\section{Orthogonality of Brillouin grating}

\begin{figure}
\centering
\includegraphics[width=.85\textwidth]{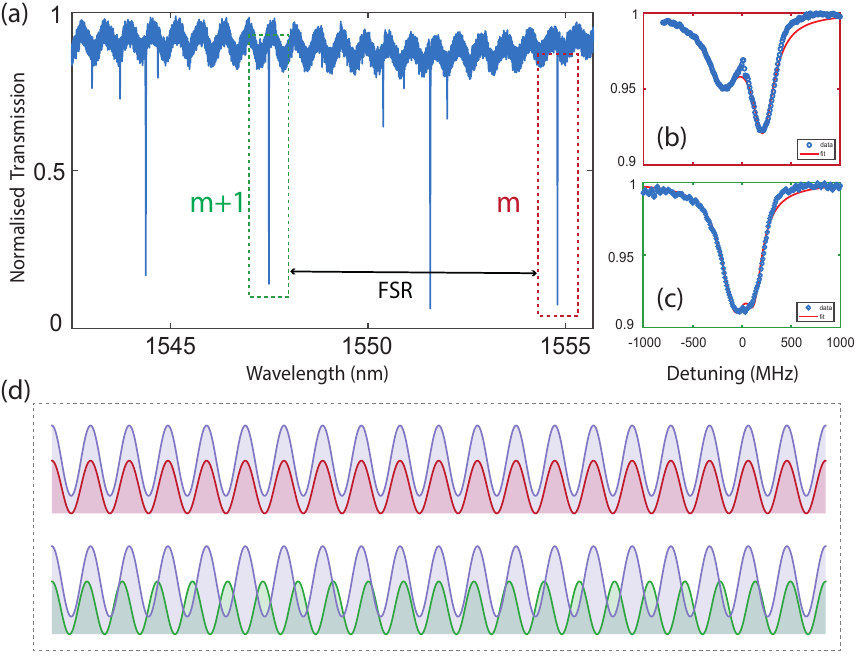}%
\caption{a) Normalized experimental cavity transmission spectrum, showing the optical mode used in the experiments ($m_{\mathrm{opt}}=186$) near 1555 nm (dashed red box), along with the next azimuthal order WGM ($m_{\mathrm{opt}}+1=187$) separated in wavelength by an FSR (dashed green box). In the presence of the strong refractive index grating created by pumping mode $m_{\mathrm{opt}}$ above  the Brillouin lasing threshold, the pumped mode measured by the probe laser reveals strong mechanically induced optical coupling (b), while the adjacent $m_{\mathrm{opt}}+1$  optical mode  remains unsplit  (c).  d) Schematic illustration of mode orthogonality between mode $m_{\mathrm{opt}}=10$ (red) and its associated refractive index modulation (blue) and mode $m_{\mathrm{opt}}+1=11$ (green).
}%
\label{fig:orthogonality}%
\end{figure}

The whispering gallery modes constitute orthogonal eigenmodes of the electromagnetic field confined inside the optical resonator. Because of this orthogonality,  the superfluid surface deformation (and its associated  refractive index modulation) caused by driving one WGM  should in principle leave other WGMs unaffected.
This is indeed what we observe in the experiments. Figure \ref{fig:orthogonality}(a) shows a transmission spectrum of our microresonator, with the WGM used in the experiments ($m_{\mathrm{opt}}=186$) highlighted in red. The WGM of the same mode family with the next azimuthal order ($m_{\mathrm{opt}}+1=187$), separated from the first by a free spectral range, is highlighted in green. Using the pump-probe setup described in the main text and section  \ref{sectionopticalstrongcoupling}, we use the strong pump tuned to the $m_{\mathrm{opt}}$ WGM to initiate Brillouin lasing. In this regime, the weak probe scanned over the $m_{\mathrm{opt}}$ WGM reveals a doublet splitting (Fig. \ref{fig:orthogonality}(b)), a manifestation of the strong coupling between forward and backward propagating optical fields  mediated by the superfluid index grating. In contrast, sweeping over the adjacent $m_{\mathrm{opt}}+1$ WGM in the same lasing regime shows this mode remains unaffected and maintains its Lorentzian shape, see Fig. \ref{fig:orthogonality}(c).

A schematic illustration of this orthogonality is shown in Fig. \ref{fig:orthogonality}(d), with lower azimuthal orders plotted here for clarity. The top panel shows the intensity profile, proportional to $E^2$, of the $m_{\mathrm{opt}}=10$ WGM along the circumference of the resonator (red), along with the associated superfluid film deformation generated by the Brillouin scattering process (blue). The lower panel shows this same surface deformation (blue) along with the field intensity of the $m_{\mathrm{opt}}+1=11$ WGM.
Because of the WGM orthogonality, the refractive index modulation created by mode $m_{\mathrm{opt}}$ leads to net change in optical path length for the $m_{\mathrm{opt}}+1$ WGM, and hence no energy shift (Eq.(\ref{EqDeltaE1})) and a zero $g_{0,\mathrm{rp}}$ and $g_{opt}$, see section \ref{subsecequationsofmotiongopt}.

\section{Analytical theory - non-depleted pump regime}
\label{sectionanalyticaltheory}

\subsection{Hamiltonian formalism}

Here we detail the analytical Brillouin scattering theory, which describes the superfluid Brillouin system in the non-depleted pump regime, an approximation valid below the lasing threshold. The general interaction Hamiltonian is determined by the overlap integral of the optical and acoustic fields~\cite{Tomes2011,baker_theoretical_2016}. We derive it below, based on a perturbation theory approach. The energy shift $\Delta E$ experienced by the optical field due to the presence of the superfluid is given by:
\begin{equation}
\Delta E = \int _{r=0}^R \int_{\theta=0}^{2\pi} \int_{z=0}^{d_0+\eta\left( r, \theta \right)}\, \frac{1}{2}\, \varepsilon_0\, \left(\varepsilon_{\mathrm{sf}}-1\right) |E|^2\,  r\,\mathrm{d}r \, \mathrm{d}\theta\, \mathrm{d}z,
\label{EqDeltaE1}
\end{equation}
where $\varepsilon_0=8.85\times 10^{-12}$ F/m is the vacuum permittivity,
$\varepsilon_{\mathrm{sf}}$ is the relative permittivity of superfluid helium, $\eta$ is the out-of-plane displacement of the superfluid surface beyond a mean height $d_0$ due to the  acoustic wave (third sound), $E$ is the WGM electric field and $R$ is the radius of the disk resonator. Given the film thickness is only a few nanometres, the electric field can be treated as constant over the height of the film. Thus the volume integral Eq. (\ref{EqDeltaE1}) can be reduced to an integral over the surface of the disk resonator:
\begin{equation}
\Delta E = \int _{0}^R \int_{0}^{2\pi} \, \frac{1}{2} \, \varepsilon_0\, \left(\varepsilon_r-1 \right) \left( d_0+\eta\left( r,\theta \right) \right)|E|^2\,  r\,\mathrm{d}r \, \mathrm{d}\theta\, 
\end{equation}
We rewrite this expression in terms of photon and phonon annihilation and creation operators:
\begin{equation}
\Delta E = \int_\mathrm{{surface}} \, \frac{1}{2} \, \varepsilon_0 \, \left(\varepsilon_r - 1 \right)  \left[ \underbrace{\Psi_b(\vec{r})\left(b_q + b_q^{\dagger} + b_{-q} + b_{-q}^{\dagger}\right)}_{\mathrm{AC}}+\underbrace{d_0}_{\mathrm{DC}}\right] \left[\vert\Psi_a(\vec{r})\vert^2 |a_k +a_{-k}|^2\right]\,\mathrm{d}A.
\label{EqDeltaE3}
\end{equation}
Here $\Psi_a(\vec{r})=E\left( \vec{r} \right) \sqrt{\frac{\hbar \omega}{\int \frac{1}{2} \, \varepsilon_0 \,\varepsilon_r \, E^2\, \mathrm{d}V}}$ is the electric amplitude per photon, $\Psi_b(\vec{r})$ is the acoustic amplitude in the ground state, $a_k$ ($b_q$) and $a_{-k}$ ($b_{-q}$) are respectively the photon (phonon) annihilation operators acting on the forward and backward propagating optical (acoustic) modes.   The subscripts $k$ ($q$) refer respectively to the wavenumbers of the optical (acoustic) wave and the signs in front of $k$ and $q$ indicate the propagation direction, with the convention that the optical pump travels in the positive direction.
In the limit that the Brillouin frequency is much smaller than the optical frequency ($\ob\ll \omega_{\mathrm{opt}}$), momentum and energy conservations require that $q = 2k$ 
 for the Brillouin process.
 
The acoustic term in Eq. (\ref{EqDeltaE3}) can be broken into two components: a DC  component proportional to the mean film thickness $d_0$ which gives the DC shift of the optical resonance frequency due to the superfluid helium film covering the resonator, and an AC component describing the interaction of the surface acoustic wave with the intracavity optical field. Neglecting the DC part, the interaction Hamiltonian takes the form:
\begin{equation}
\begin{split}
H_\mathrm{{int}} &=- \int _\mathrm{{surface}} \, \frac{1}{2} \, \varepsilon_0 \,  \left(\varepsilon_r-1 \right) \Psi_b(\vec{r}) \vert\Psi_a(\vec{r})\vert^2\, \, \mathrm{d} A \,(b_q + b_q^{\dagger} + b_{-q} + b_{-q}^{\dagger})|a_k + a_{-k} |^2 \\
&= -\hbar \,g_{0,\mathrm{rp}} \,(b_q + b_q^{\dagger} + b_{-q} + b_{-q}^{\dagger})( a_k^{\dagger}+ a_{-k}^{\dagger})(a_k + a_{-k}),\\
\end{split}
\end{equation}
where $g_{0,rp}$ is the single photon optomechanical coupling rate from radiation pressure \cite{Aspelmeyer2014,baker_theoretical_2016}:
\begin{equation}
g_{0,\mathrm{rp}}=\int_{\mathrm{surface}} \, \frac{1}{2}\, \varepsilon_0 \left(\varepsilon_{\mathrm{sf}} -1 \right) \, \Psi_b(\vec{r}) \, \vert\Psi_a(\vec{r})\vert^2\, \, \mathrm{d} A 
\end{equation}
The interaction Hamiltonian is further reduced by energy and momentum conservation arguments to the following form:
\begin{equation}
H_\mathrm{{int}} = -\hbar\, g_{0,\mathrm{rp}}( b_q^{\dagger} a_k a_{-k}^{\dagger} +   b_{-q} a_k a_{-k}^{\dagger}+ b_{-q}^{\dagger} a_k^{\dagger}  a_{-k} + b_q a_k^{\dagger} a_{-k}  )
\label{4terms}%
\end{equation}
The first two terms correspond respectively to the Stokes and anti-Stokes scattering process for the forward propagating optical field $a_k$, while the third and fourth terms correspond respectively to the Stokes and anti-Stokes process acting on the counter-propagating optical field $a_{-k}$. 

Note that in many other works~\cite{ObservationOfSpontaneous,HighFrequency}, the resonator's optical spectrum is engineered such that pump and Stokes fields are resonant with two distinct optical modes separated by the Brillouin shift. Compared with the pump and Stokes, the non-resonant anti-Stokes field experiences a very low optical density of states and can therefore be neglected. Here, because of the small Brillouin shift ($\sim6$ MHz), we must keep both the Stokes and anti-Stokes terms. In addition, the frequencies of the two counter-propagating optical modes are degenerate, such that the frequencies of $a_k$ and $a_{-k}$ are both treated as $\Delta$ in the full Hamiltonian description in a frame rotating at the laser frequency $\omega_L$, where $\Delta$ is defined as $\omega_L - \omega_k$:
\begin{equation}
\label{full4}
H = \underbrace{-\hbar\Delta a_k^{\dagger}a_k - \hbar\Delta a_{-k}^{\dagger} a_{-k}}_{\text{optical}} + \underbrace{\hbar\ob b_q^{\dagger} b_q + \hbar\ob b_{-q}^{\dagger} b_{-q}}_{\text{mechanical}} \underbrace{- \hbar g_{0,\mathrm{rp}} (b_q a_k^{\dagger} a_{-k} + b_{-q}^{\dagger} a_k^{\dagger} a_{-k} + b_{-q} a_{-k}^{\dagger} a_k + b_q^{\dagger} a_{-k}^{\dagger} a _k)}_{\text{Brillouin interaction}} 
\end{equation}

\subsection{Equations of motion}

From this Hamiltonian we derive the equations of motion. Including the drive $a_{in}$ of the pump field, the fountain pressure coupling ($g_{0,\mathrm{fp}} = g_{0,\mathrm{tot}} - g_{0,\mathrm{rp}}$), the native backscattering $\kappa_b$, the thermal drive $b_{in}$ ($b_{-in}$) of the acoustic fields $b_q$ ($b_{-q}$) respectively~\cite{HighFrequency,bowen2015quantum} and dissipation we obtain:
\begin{equation}
\label{full41}
\dot{a}_k = i\Delta a_k -\frac{\kappa}{2} a_k + i g_{0,\mathrm{rp}} (b_q a_{-k} + b_{-q}^{\dagger} a_{-k}) + i\kappa_b a_{-k} + \sqrt{\kappa_{ext}}\, a_{in}
\end{equation}
\begin{equation}
\label{full42}
\dot{a}_{-k} = i\Delta a_{-k} - \frac{\kappa}{2} a_{-k} + i g_{0,\mathrm{rp}} (b_{-q}a_k + b_q^{\dagger}a_k)+ i\kappa_b a_{k}
\end{equation}
\begin{equation}
\label{full43}
\dot{b}_q = -i \ob b_q - \frac{\Gamma}{2}b_q + i g_{0,\mathrm{tot}}\, a_{-k}^{\dagger} a_k + \sqrt{\Gamma} \,b_{in}(t)
\end{equation}
\begin{equation}
\label{full44}
\dot{b}_{-q} = -i \ob b_{-q} - \frac{\Gamma}{2} b_{-q} + i g_{0,\mathrm{tot}}\, a_{k}^{\dagger} a_{-k} + \sqrt{\Gamma} \,b_{-in} (t),
\end{equation}
where $\kappa$ is the optical decay rate, $\kappa_{ext}$ the coupling rate of the optical cavity to the tapered fibre, $n_q$ ($n_{-q}$) the thermal occupation of the co-propagating (counter-propagating) acoustic field, $\Gamma$ the intrinsic mechanical damping rate, and $b_{in}(t)$ the thermal drive, which obeys the Markovian noise process such that
$
\label{thermal1}
\langle b_{in}^{\dagger}(t)b_{in}(t')\rangle = n_{q}\,\delta (t-t')
$
 and
$
\label{thermal2}
\langle b_{in}(t)b_{in}^{\dagger}(t') \rangle = (n_{q}+1)\delta (t-t')
$~\cite{bowen2015quantum}.
Note that light can drive the acoustic wave through both the radiation pressure and fountain pressure interactions, but the acoustic wave affects light only through thickness fluctuations, namely dispersive coupling mediated by radiation pressure.

We make the assumption that $a_k$ is a non-depleted pump, and the native backscattered light is also not depleted. Then the equations of motion can be linearized by decomposing $a_k$ and $a_{-k}$ into an average coherent amplitude $\alpha_k$ (or $\alpha_{-k}$) given by the steady state solution, and a time-variant part $\delta a_k$ (or $\delta a_{-k}$) describing the dynamics of the system. First we neglect the Brillouin interaction and solve the steady-state of the pump and the backscattered light from Eq.~\eqref{full41} and Eq.~\eqref{full42}, and obtain:
\begin{equation}
\alpha _k = \sqrt{n_{cav}}\,e^{-i\omega_L t} = {\frac{\sqrt{\kappa_{ext}}\,\alpha_{in}}{-i\Delta + \kappa/2 + \frac{\kappa_b^2}{-i\Delta + \kappa/2}}},
\end{equation}
\begin{equation}
\alpha_{-k}=\frac{i\kappa_b }{-i\Delta + \kappa/2}\alpha_k.
\end{equation}
Here we define $n_{cav,k}$ and $n_{cav,-k}$ as the intracavity photon number approximated by the steady state solution of the pump and backscattered light respectively: $n_{cav,k} = \abs{\alpha_k}^2$, $n_{cav,-k} = \abs{\alpha_{-k}}^2$.
%

It is important to point out here, that the presence of weak backscattering manifests as a suppression of Brillouin amplification. This suppression arises primarily from two mechanisms: first, through the depletion of pump photons from backscattering, and second, those backscattered photons (now counter-propagating) actively cool the acoustic mode that is being amplified by the pump. Nevertheless, the lasing threshold behavior still remains qualitatively the same, albeit occuring at higher injected optical powers. To concisely describe the lasing threshold behavior the following sub-sections (Sec.\ref{Brillouin_Lasing_Threshold} and Sec.\ref{Backreflected_light_power_spectrum}) will neglect the effect of backscattering. However, when processing the experimental data to extract the total optomechanical coupling rate the full dynamical equations will be considered (see Sec.\ref{sectionBrillouinODE}).   

Subtracting the steady state solutions from Eqs.(\ref{full41},\ref{full42},\ref{full43},\ref{full44}) and neglecting backscattering, optical vacuum noises and higher order terms of the time-variant variables, the Fourier transformed equations of motion are: 
\begin{equation}
-i\omega \delta a_{k}(\omega) = i\Delta \delta a_{k}(\omega) - \frac{\kappa}{2}\delta a_{k}(\omega) + ig_{\mathrm{0,rp}}\alpha_{-k}( b_{q}(\omega) + b_{-q}^{\dagger}(\omega)), \label{full4111}
\end{equation}
\begin{equation}
-i\omega\delta a_{-k}(\omega) = i\Delta\delta a_{-k}(\omega) - \frac{\kappa}{2}\delta a_{-k}(\omega) + ig_{\mathrm{0,rp}}\alpha_{k}( b_{-q}(\omega) + b_q^{\dagger}(\omega)), \label{full4222}
\end{equation}
\begin{equation}
-i\omega b_q(\omega) = -i\ob b_q(\omega) - \frac{\Gamma}{2}b_q(\omega) + ig_{\mathrm{0,tot}}(\alpha^*_{-k}\delta a_k(\omega) + \alpha_{k}\delta a_{-k}^{\dagger}(\omega)) + \sqrt{\Gamma} \, b_{in}(\omega),\label{full4333}
\end{equation}
\begin{equation}
-i\omega b_{-q}(\omega) = -i\ob b_{-q}(\omega) - \frac{\Gamma}{2} b_{-q}(\omega) + ig_{\mathrm{0,tot}}(\alpha^*_k \delta a_{-k}(\omega) + \alpha_{-k}\delta a^\dagger_{k}(\omega)) + \sqrt{\Gamma} \, b_{-in}(\omega). \label{full4444}
\end{equation}
In the Fourier domain, the temporal correlation of the thermal noise becomes $\frac{1}{2\pi}\int^{\infty}_{-\infty}\langle b^\dagger_{in}(\omega)b_{in}(\omega')\rangle d\omega' \,=\,n_q$, where $b_{in}^\dagger(\omega)$ is the Fourier transform of the complex conjugate of $b_{in}(t)$.

\subsection{Brillouin lasing threshold}
\label{Brillouin_Lasing_Threshold}
Below the Brillouin lasing threshold of the co-propagating acoustic wave $b_q$,  the counter-propagating acoustic wave $b_{-q}$ will be damped out.
We therefore neglect here the counter-propagating wave $b_{-q}$ in the equations of motion, a valid approximation to predict the dynamics of the forward travelling wave $b_q$ below the lasing threshold. Neglecting the counter-progating wave, the equations of motion are solved using the identity $b^{\dagger}(\omega) = [b(-\omega)]^{\dagger}$, and the co-propagating acoustic wave amplitude is:
\begin{equation}
b_q(\omega) = \frac{\sqrt{\Gamma} \, b_{in}(\omega)}{-i(\omega -\ob) + \Gamma/2 - \frac{\abs{g_{\mathrm{0,rp}}g_{\mathrm{0,tot}}}n_{cav,k}}{-i\omega+i\Delta +\kappa/2}}
\label{bqsimple}
\end{equation}
We identify the third term in the denominator of Eq.(\ref{bqsimple}) as:
\begin{equation}
\Sigma_{b_q}(\omega)  = \frac{-\abs{g_{\mathrm{0,rp}}g_{\mathrm{0,tot}}}n_{cav,k}}{-i\omega+i\Delta +\kappa/2} = i\delta\ob(\omega) + \frac{\Gamma_{\mathrm{opt}}(\omega)}{2}
\label{Sigmabq}
\end{equation}
The imaginary part of $\Sigma_{b_q}(\omega)$ leads to a shift in the Brillouin frequency $\delta \ob$ (optical spring effect), while its real part leads to a modified effective mechanical damping rate $\Gamma_{\mathrm{eff}}$ through the addition of an optical damping term $\Gamma_{\mathrm{opt}}$.
In our system, the total optical cavity linewidth ($\kappa/2\pi=284$ MHz) is much larger than the acoustic frequency ($\ob/2\pi=7.3$ MHz) 
and the pump light is close to resonance ($(\Delta-\omega)\ll \kappa$). As a result, $\Sigma_{b_q}\simeq \Re{\Sigma_{b_q}}$ and primarily contributes to changing the effective mechanical linewidth through $\Gamma_{\mathrm{opt}}\simeq-\frac{4\abs{g_{\mathrm{0,rp}}g_{\mathrm{0,tot}}}(n_{cav,k} - n_{cav,-k})}{\kappa_{\mathrm{eff}}}$. 
The effective mechanical linewidth $\Gamma_{\mathrm{eff}}$ of the co-propagating acoustic wave can thus be expressed as: 
\begin{equation}
\Gamma_{\mathrm{eff}}=\Gamma - \frac{4\abs{g_{\mathrm{0,rp}}g_{\mathrm{0,tot}}}}{\kappa}\zeta n_{cav,k}
\label{Gammaeff2}
\end{equation}
where the new unitless term, $\zeta$, is added to account for the suppression of linewidth narrowing from backscattered photons. The value of this term is determined via numerical techniques in Sec.\ref{sectionBrillouinODE}. 

\begin{figure}
\centering
\includegraphics[width=0.5\textwidth]{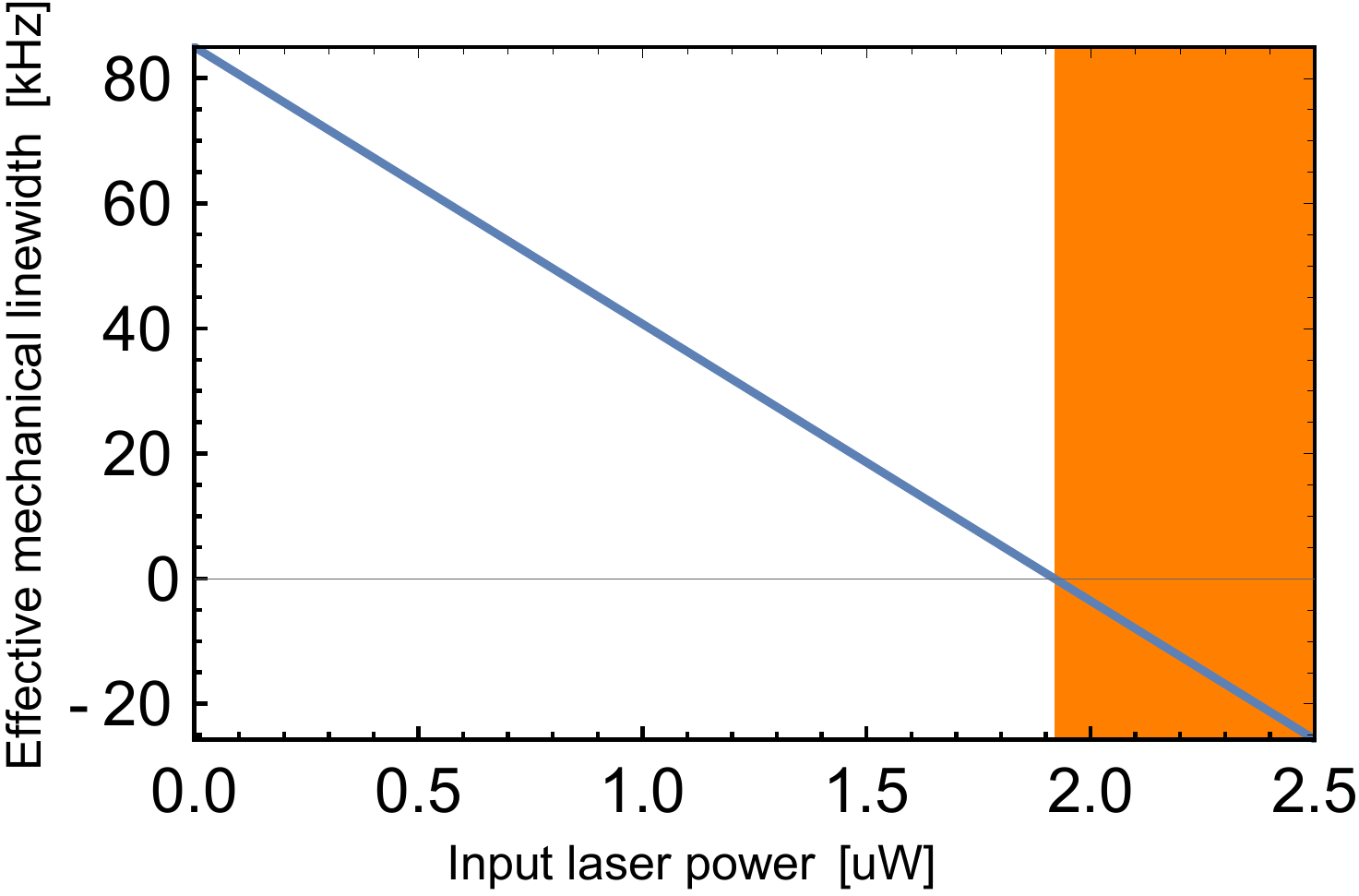}
\caption{Effective mechanical linewidth $\Gamma_{\mathrm{eff}}$ of the forward propagating acoustic wave versus input laser power (blue line), plotted using Eq.~(\ref{Gammaeff2}). The orange shading corresponds to the region above the Brillouin lasing threshold, for which the analytical theory is no longer valid (see section \ref{sectionBrillouinODE} for above threshold analysis).  }
\label{fig:GammaBq}
\end{figure}

Figure \ref{fig:GammaBq} plots $\Gamma_{\mathrm{eff}}$ as a function of input laser power, with the following parameters: acoustic resonance frequency $\ob/2\pi$ $= 7.3$ MHz, laser (pump) frequency $w_L/2\pi = 193$ THz, fiber-to-cavity coupling rate $\kappa_{ext}=180$ MHz, optical decay rate $\kappa/2\pi = 284$ MHz, the radiation pressure optomechanical coupling rate $g_{0,\mathrm{rp}}$ is estimated to be $11$ kHz in Section~\ref{sectionestimationofg0} of this supplementary material, optical detuning $\Delta = \ob$,  intrinsic mechanical linewidth $\Gamma\,=\, 85$ kHz, total optomechanical coupling $g_{0,\mathrm{tot}}\,=\, 133$ kHz and $\zeta=0.22$. The effective linewidth approaches zero at 1.8 $\mu$W input power, which corresponds to the Brillouin lasing threshold. 

Note as mentioned earlier that this analytical result is only valid in the non-depleted pump regime, and only describes the behaviour below the threshold. The above threshold behaviour is explained by numerically solving the full equations of motion without the non-depleted pump approximation, which accounts for the saturation of the Stokes amplitude (see section \ref{sectionBrillouinODE} for more details).

\subsection{Backreflected light power spectrum}
\label{Backreflected_light_power_spectrum}
\begin{figure}[t!]
\centering
\includegraphics[width=0.6\textwidth]{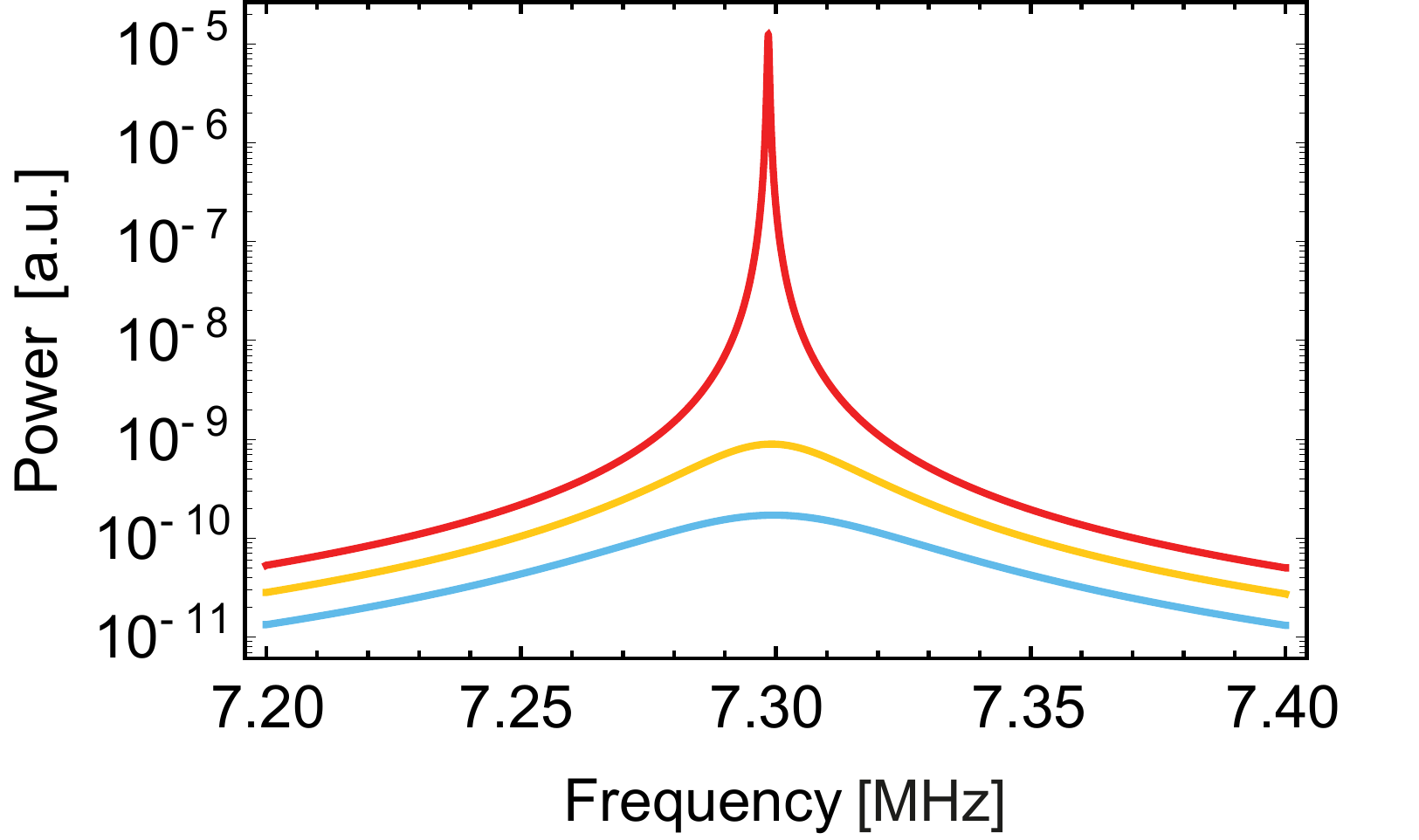}
\caption{Analytical calculation of the reflected light power spectrum in the non-depleted pump regime, showing the lasing of the Stokes sideband. Curves corrspond to $P_{in}=0.5\, \mu$W (blue); $P_{in}=1\, \mu$W (yellow) and $P_{in}=1.8\, \mu$W (red - at phonon lasing threshold).}
\label{fig:SaMkaMk}
\end{figure}

We solve here for the backscattered light power spectrum, used for the solid fit in Fig. 2 of the main text.
Following the same approach used to determine $b_q$, we obtain the backreflected light spectral amplitude $\delta a_{-k}(\omega)$:
\begin{equation}
\delta a_{-k}(\omega) = \frac{ig_{\mathrm{0,rp}}\alpha_k\,b^{\dagger}_q(\omega) }{-i\omega -i\Delta + \kappa/2}
\label{aMk0}
\end{equation}
The power spectrum of the reflected light then takes the following form using the Wiener-Khinchin theorem:
\begin{equation}
\begin{split}
S_{\delta a_{-k}\delta a_{-k}}(\omega) &= \frac{1}{2\pi}\int_{-\infty}^{\infty}d\omega'\big\langle [\delta a_{-k}(-\omega)]^{\dagger} \delta a_{-k}(\omega ') \big\rangle \\
 &= \frac{1}{2\pi}\int_{-\infty}^{\infty}d\omega' \Big \langle \frac{- ig_{\mathrm{0,rp}}\alpha^*_k \, b_q(\omega)}{-i\omega + i\Delta + \kappa/2} \cdot \frac{ig_{\mathrm{0,rp}}\alpha_k\, b^{\dagger}_q (\omega ')}{-i\omega' - i\Delta + \kappa/2} \Big \rangle
\end{split}
\label{eq:SaMkaMk1}
\end{equation}
After substituting the solution of $b_q$ given by Eq.(\ref{bqsimple}) into Eq.(\ref{aMk0}), the explicit solution of $\delta a_{-k}(\omega)$ is:
\begin{equation}
\delta a_{-k}(\omega) = \frac{ig_{\mathrm{0,rp}}\alpha_k}{-i\omega -i\Delta + \kappa/2}\cdot  \Bigg( \frac{\sqrt{\Gamma}b^{\dagger}_{in}(\omega)}{-i\omega -i\ob + \Gamma/2 - \frac{\abs{g_{\mathrm{0,rp}}g_{\mathrm{0,tot}}}n_{cav,k}}{-i\omega-i\Delta +\kappa/2}} \Bigg),
\end{equation}
where we recognize both an optical cavity and acoustic response components. Defining the cavity susceptibility as $\chi_{-k}(\omega)$ and the Brillouin-interaction modified acoustic susceptibility as $\chi_q(\omega)$, with:
\begin{equation}
\label{CA}
\begin{split}
\chi_{-k}^{-1}(\omega) &=-i\omega -i\Delta + \kappa/2  \\
\chi_q^{-1}(\omega) &=  -i(\omega -\ob) + \Gamma/2 - \frac{\abs{g_{\mathrm{0,rp}}g_{\mathrm{0,tot}}}n_{cav,k}}{-i\omega+i\Delta +\kappa/2} ,\\
\end{split}
\end{equation}
we rewrite the expression of $\delta a_{-k}$ under the more convenient form: \\$\delta a_{-k}(\omega) = ig_{\mathrm{0,rp}}\alpha_k\,\chi_{-k}(\omega) [\sqrt{\Gamma}b^{\dagger}_{in}(\omega)\chi_q^*(-\omega)]$. The reflected light power spectrum is then reduced to:
\begin{equation}
\label{eq:SaMkaMk_q}
S_{\delta a_{-k}\delta a_{-k}}(\omega) = g_{\mathrm{0,rp}}^2n_{cav,k}\Gamma\abs{\chi_{-k}(\omega)}^2 \left[\abs{\chi_q^*(-\omega)}^2\cdot (n_q +1)\right]
\end{equation}
When the pump power is increased, the co-propagating acoustic wave is increasingly amplified. The Stokes sideband generated by the co-propagating wave is plotted in Fig.~\ref{fig:SaMkaMk}, using the experimental parameters specified above. The power spectra show how the pump field drives the co-propagating acoustic wave, eventually resulting in lasing around 1.8 $\mu$W.

\subsection{Bulk heating from the pump beam}
As discussed in the main text, the fountain pressure contribution to the optical forcing arises from superfluid flow induced from optical heating. Given this, one might expect the thermal bath experienced by the acoustic wave to strongly depend on input power. However, the strength of these two effects  (i.e. bulk heating and increased forcing) can vary independently, according to material properties and the overlap of the `hot-spot' to the mechanical mode profile~\cite{Metzger_Nat_04,Jourdan_PRL_08}. Indeed, it has been shown that such thermal effects do not necessarily preclude ground state cooling~\cite{Restrepo_CRP_2011}.

\begin{figure}
\centering
\includegraphics[width=0.6\textwidth]{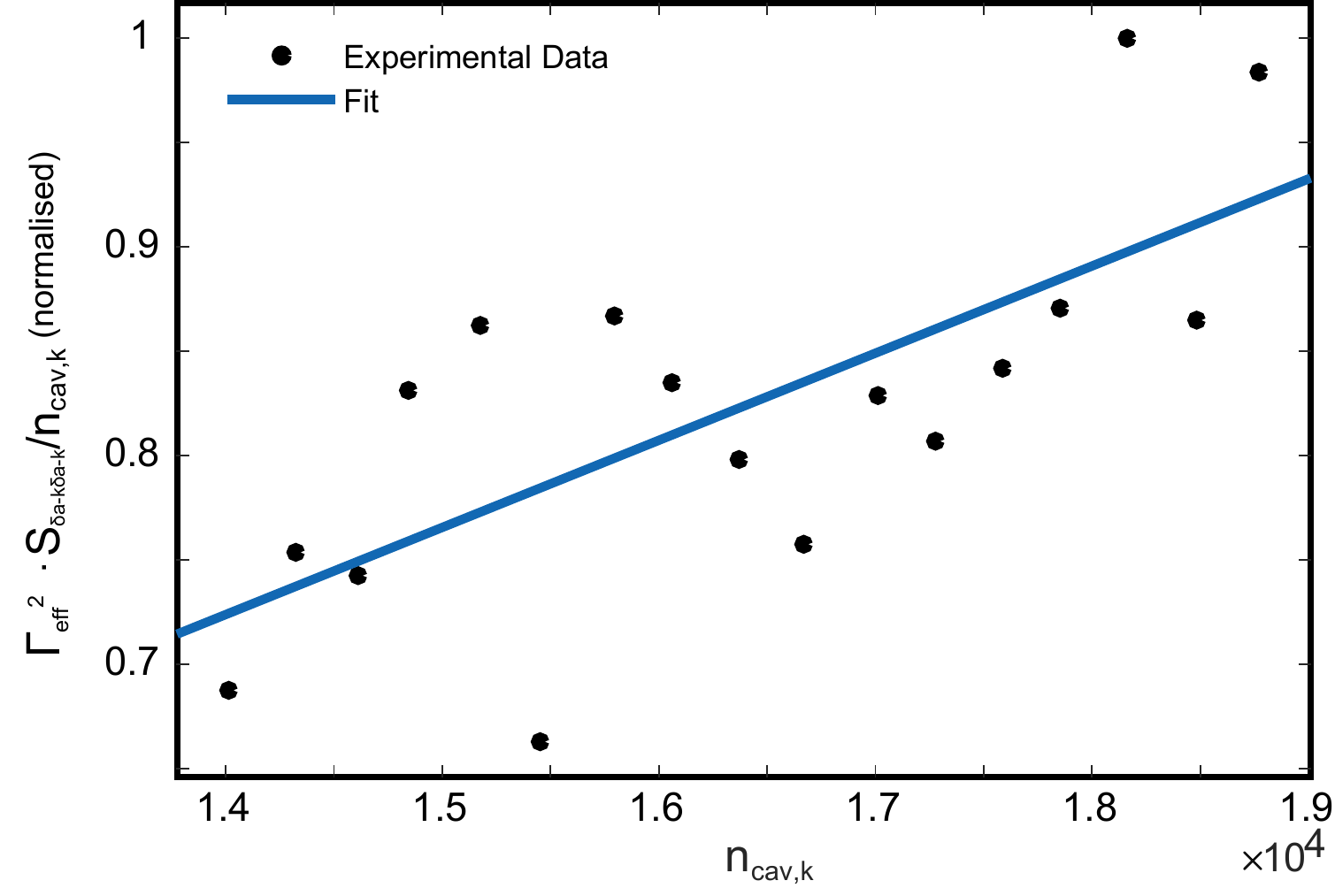}%
\caption{Experimentally observed increase of the thermal bath occupation due to optical absorption in silica. The vertical axis is normalized, and represents the right hand side of Eq.~\ref{eq:heating_2}. The ratio of the slope to the intercept gives $A/n_q$.}%
\label{fig:passive_heating}%
\end{figure}

In the absence of bulk heating, the bath occupancy is related to the Stokes power spectrum and the effective acoustic linewidth. 
This can be seen by neglecting quantum noise terms and rearranging Eq.~\ref{eq:SaMkaMk_q} with $\omega=\ob$,

\begin{equation}
n_q  \propto \frac{\Gamma^2_{\mathrm{eff}}S_{\delta a_{-k}\delta a_{-k}}(\ob)}{n_{cav,k}}
\label{eq:heating_1}
\end{equation}
This equation shows that, for all pump powers, the RHS of Eq.~\ref{eq:heating_1} is constant and related to the bath occupation through a proportionality constant. However, in presence of optical absorption in the silica disk the thermal bath occupancy experienced by the acoustic mode will increase with pump power. Including this passive heating, Eq.~\ref{eq:heating_1} becomes:

\begin{equation}
n_q + A\, n_{\mathrm{cav,k}} \propto \frac{\Gamma^2_{\mathrm{eff}}S_{\delta a_{-k}\delta a_{-k}}(\ob)}{n_{\mathrm{cav,k}}},
\label{eq:heating_2}
\end{equation}
where $n_{\mathrm{cav,k}}$ is the intracavity photon number, and $A$ quantifies the conversion from optical photons to acoustic phonons. The product of experimentally measured parameters given on the RHS of Eq.~\ref{eq:heating_2} is plotted in Fig.~\ref{fig:passive_heating}. Conveniently, the ratio of the slope to the intercept of the fit is equal to $A/n_q$, regardless of the measurement efficiency and measurement gains. At the fridge's base temperature ($\sim20$ mK), with zero input power, the thermal bath occupancy is $n_q\sim 40$. Therefore, from the fitting in Fig.~\ref{fig:passive_heating}, the conversion from optical photons to acoustic phonons is estimated to be $\sim0.02$\%. 
For example, just below lasing threshold, at 1.7 $\mu$W we have $\sim2 \times 10^4$ intracavity photons, which will contribute $\approx 400$ phonons to the thermal bath. 

\section{Full numerical solving of Brillouin equations of motion}
\label{sectionBrillouinODE}

Fig. 3 in the main text shows Brillouin lasing as manifested in (a) the fast rise, and subsequent saturation (above \SI{1.8}{\muw}), of the Stokes sideband amplitude with increasing input power, and (b) the power spectral density of the back-reflected light for \SI{5.1}{\muw} input power. We illustrate these observations further here by showcasing the time dynamics associated with the spectrum in Fig. 3(b) of the main text. The spectrum and the time-dynamics presented in this section are obtained through numerical integration of the dynamical differential equations of motion (Eqs. \ref{full41}-\ref{full44}). The analytical expression for the effective linewidth of the Stokes sideband $\Gamma_{\text{eff}}$ (Eq. (\ref{Gammaeff2}), or Eq. (1) in the main text) is valid only in the undepleted pump regime. While this provides an estimate of the power at which the effective mechanical damping $\Gamma_{\text{eff}}\rightarrow 0$ (the lasing threshold), it breaks down for input powers exceeding the threshold. Numerical analysis is therefore needed to provide quantitative information in the above-threshold regime. The following system is numerically integrated:

\begin{align}
\label{eq:NUMfull1}
\dot{a}_k &= i\Delta a_k -\frac{\kappa}{2} a_k + i g_{0,\mathrm{rp}} (b_q + b_{-q}^{\dagger} ) a_{-k} +i \kappa_b a_{-k} + \sqrt{\kappa_{ext}} a_{in} \\
\dot{a}_{-k} &= i\Delta a_{-k} - \frac{\kappa}{2} a_{-k}+ i g_{0,\mathrm{rp}}(b_{-q}+ b_q^{\dagger}) a_k + i \kappa_b a_{k} \\
\dot{b}_q &= -i \ob b_q - \frac{\Gamma}{2}b_q + i g_{0,\mathrm{tot}} a_{-k}^{\dagger} a_k  \\
\dot{b}_{-q} &= -i \ob b_{-q} - \frac{\Gamma}{2} b_{-q} + i g_{0,\mathrm{tot}} a_{k}^{\dagger} a_{-k} 
\end{align}
The simulation parameters and their numerical values are summarized in table \ref{tab:parameters}. With respect to the system Eqs. \ref{full41}-\ref{full44}, the thermal noise terms $\sqrt{\Gamma}\,b_{-in} (t)$ and $\sqrt{\Gamma}\,b_{in}(t)$ are omitted. While the presence of thermal noise is required in order to seed the lasing process, it can be replaced by non-zero starting conditions. The initialization for the presented simulation was $a_k=a_{-k}=b_q=0$, and $b_{-q}=400$. Some initial population (here $|b_{-q}|^2 = 16 \times 10^4$ phonons) in the direction opposite to the pump photons $a_k$ then seeds the Brillouin scattering process. Indeed, the presence of thermal noise and the initialization of the system only has an impact on the very start of the simulation, not on the steady-state lasing dynamics which we are interested in here. It is worth noting here that although the population of pump photons is initialized with $|a_{k}|^2=0$, it rapidly\footnote{On a time scale $\sim 1/ \kappa \sim \mathrm{ns}$} jumps to a finite population set by the coherent laser input field $a_{in} = \sqrt{\frac{P_{in}}{\hbar \omega_L}}$. 

\begin{table}
\centering
    \begin{tabular}{l|l|l|l}
        \hline
        Parameter                                   & Symbol                           & Value  & Units         \\
           \hline
           \hline
        Acoustic damping rate                       & $\Gamma / 2 \pi $                & $85 $  & $\mathrm{kHz}$  \\ 
        Acoustic (Brillouin) frequency              & $\ob / 2 \pi $                & $6.3$    & $ \mathrm{MHz}$ \\ 
        Laser detuning                              & $\Delta$                         & $0$    & MHz               \\ 
        Overall cavity intensity decay rate         & $\kappa / 2 \pi $                & $284 $ & $\mathrm{MHz}$  \\ 
        Input coupling (external) cavity decay rate & $\kappa_{\mathrm{ext}} / 2 \pi $ & $180$  & $ \mathrm{MHz}$ \\ 
        Backscattering decay rate                   & $\kappa_{\mathrm{b}} / 2 \pi $   & $75$  & $ \mathrm{MHz}$ \\ 
        Single photon coupling strength (radiation pressure)      & $g_{\mathrm{0,rp}}/ 2 \pi $                    & $11$   & $ \mathrm{kHz}$ \\
				Single photon coupling strength (fountain pressure)            & $g_{\mathrm{0,fp}}/ 2 \pi $                    & $122$   & $ \mathrm{kHz}$ \\
        \hline
    \end{tabular}
\caption{System parameters used in numerical simulation.}
\label{tab:parameters}
\end{table}

\begin{figure}
\begin{center}
\includegraphics[scale=1.05,angle=0]{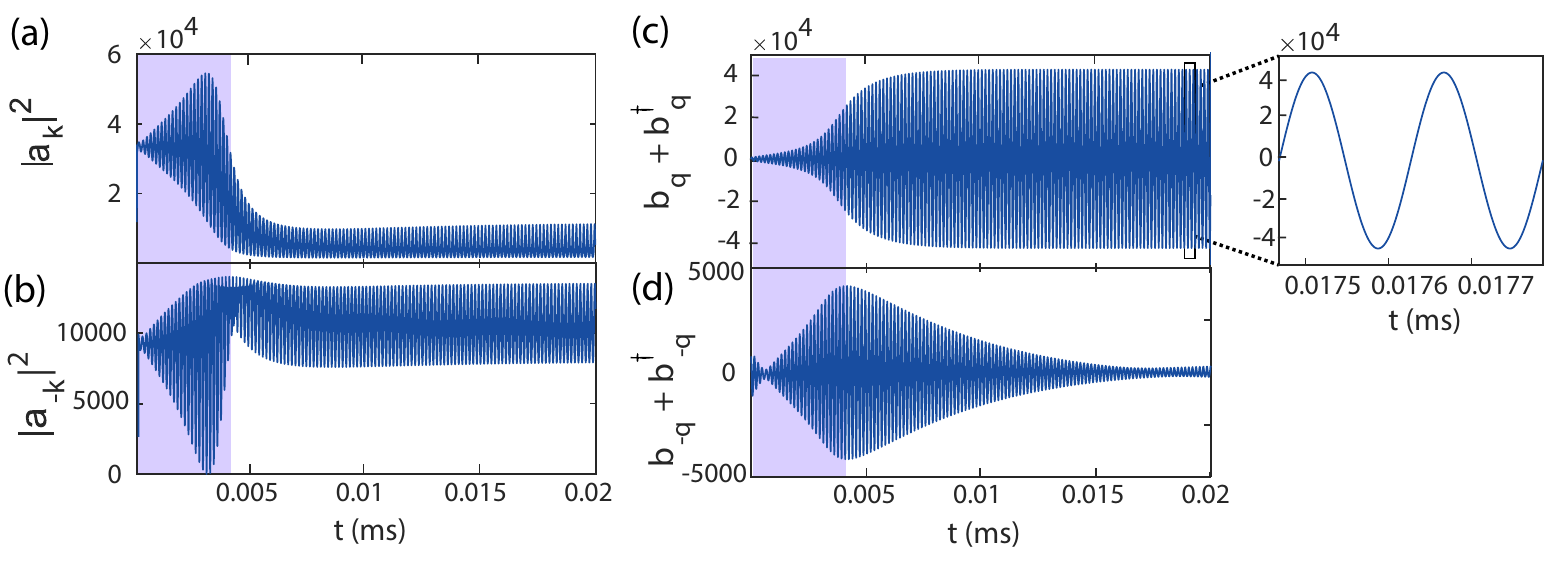}
\caption{Brillouin scattering at \SI{5.1}{\muw} input power: transient evolution towards steady-state lasing. The field amplitude is $a$, while $|a|^2$ corresponds to the intracavity photon number. $\hat{b}_{q}+\hat{b}_{q}^\dagger$ and $\hat{b}_{-q}+\hat{b}_{-q}^\dagger$ are the displacement amplitudes in units of the zero point fluctuation amplitude $x_{\mathrm{ZPF}}= \sqrt{\hbar/ 2 m \ob}$ \cite{Aspelmeyer2014}. (a) Incoming photons of amplitude $\hat{a}_{k}$ are converted to (b) backscattered photons of amplitude $\hat{a}_{-k}$ and (c) forward propagating Brillouin phonons with displacement amplitude $\hat{b}_{q}+\hat{b}_{q}^\dagger$. (d) Backpropagating acoustic wave with displacement amplitude $\hat{b}_{-q}+\hat{b}_{-q}^\dagger$. Inset: close-up of the Brillouin displacement amplitude in the lasing regime, in a time frame covering two mechanical periods ($2\times \frac{1}{\SI{6.3}{\mhz}}$). Plots (a), (b), (c) \& (d) show the transient dynamics of a longer $t= \SI{5}{\milli \second}$ simulation with $2 \times 10^7$ time steps, used to compute the power spectrum shown in Fig. \ref{fig:cascaded}. \label{fig:brillouin}}
\end{center}
\vspace{-10pt}
\end{figure}

Figure \ref{fig:brillouin} shows the evolution over time of the circulating light intensity and acoustic displacement amplitude, in the forward (a and c) and backward (b and d) directions. 
First, the forward propagating light field is dominant ($\abs{a_k}^2 > \abs{a_{-k}}^2$), and the forward propagating acoustic wave is amplified  (shaded region in Fig. \ref{fig:brillouin} c, Stokes process) by reflection of the forward travelling photons. Once this forward-travelling refractive index grating has reached a significant amplitude ($t=\sim0.005$ s), this leads to a depletion of the pump field $a_k$, and a saturation of the growth of the lasing forward Brillouin wave $b_q$. This is the mechanism behind the saturation visible in Fig. 3 (a) of the main text. After some period of transient dynamics ($t=\sim0.02$ s), the backward propagating acoustic wave (Fig. \ref{fig:brillouin} d) settles into a low amplitude steady-state.

\begin{figure}
\centering
\includegraphics[width=.6\textwidth]{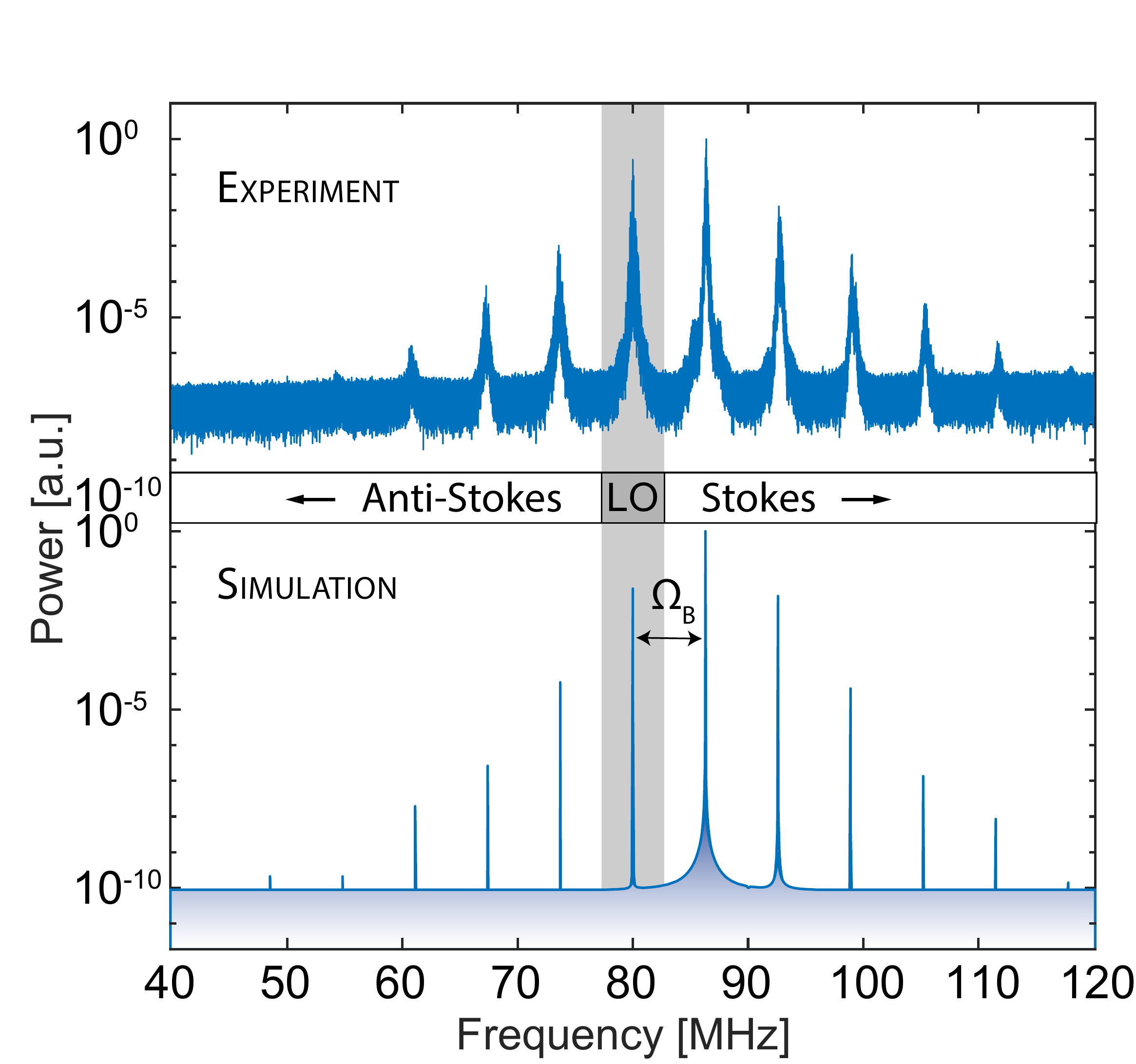}
\caption{Brillouin lasing observed in the spectrum of light reflected from the cavity at high optical powers. Top: experimental data. Bottom: $|\hat{a}_{-k}+\hat{a}_{\text{LO}}| ^2[\omega]$ obtained via numerical integration of differential equations (no free parameters). Frequencies displayed are shifted by a local oscillator beam modulated at \SI{80}{\mhz}.} 
\label{fig:cascaded}
\end{figure}

The lasing process can also be observed in the frequency domain. The power spectrum of the Brillouin-scattered light $\hat{a}_{-k}$ in Fig. \ref{fig:cascaded} (Fig. 3 (b) in main text) is obtained by taking the Fast Fourier transform over the last $\SI{4}{\milli \second}$ of the time evolution once the system has reached its steady-state. A local oscillator beam $\hat{a}_{\text{LO}}$ sinusoidally modulated at 80 MHz is added to the signal $\hat{a}_{-k}$ to explicitly mimic the experiment. Aside from the Stokes and anti-Stokes sidebands immediately on either side of the LO, a comb-like pattern is observed, with features integer multiples of the Brillouin frequency $\ob$ away from the LO. We ascribe these to non-linear sideband mixing and higher-order Stokes scattering. 
Comparing the spectral traces in Fig. \ref{fig:cascaded}, we observe very good agreement between the experimental observations and theoretical predictions through numerical simulation.

\section{Estimation of electrostrictive $g_{0,\mathrm{es}}$ of the silica disk}
\label{section_electrostriction}

\begin{figure}
\centering
\includegraphics[width=.75\textwidth]{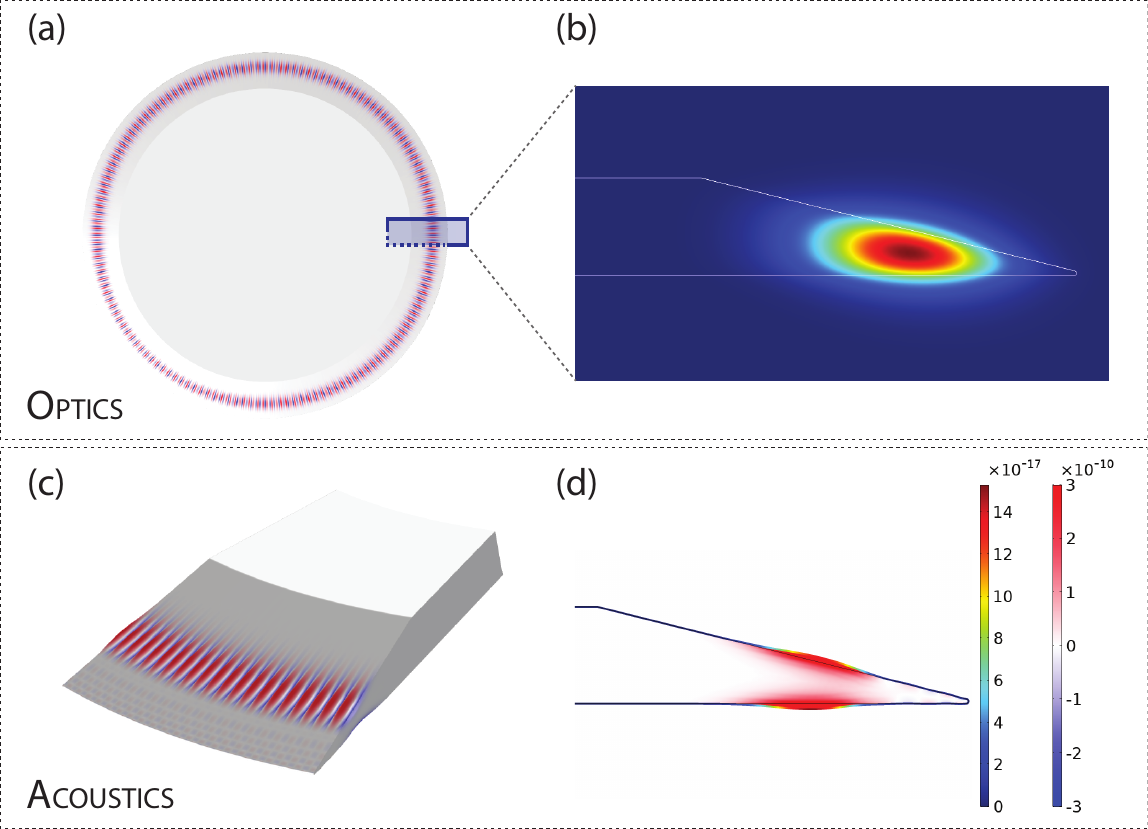}
\caption{Estimation of the electrostrictive $g_{0,\mathrm{es}}$ of the silica disk resonator. (a) Top-view of the electric field distribution for the ($p=1$; $m=186$) TE WGM of the resonator. (b) side-view of the electric field intensity distribution $\abs{E}^2$. (c) Finite element element model of a propagating acoustic wave localized in the disk wedge, with azimuthal order $m=372$ and frequency $\ob=6$ GHz. (d) Cross-section of the silica resonator displaying the zero-point surface deflection (line plot), as well as the zero-point volumetric strain (surface plot) due to the Brillouin wave, with their respective scale bars: the zero-point volumetric strain (and relative permittivity) fluctuations are on the order of $10^{-10}$ while the zero-point surface deflection is on the order of $10^{-16}$ m.} 
\label{fig:electrostrictive_g0}
\end{figure}

We compare here the magnitude of the radiation pressure coupling  between light and thickness fluctuations in the superfluid film $g_{0,\mathrm{rp}}$, to the electrostrictive coupling rate between light confined in the silica resonator and an acoustic wave propagating in the silica itself, $g_{0,\mathrm{es \,SiO2}}$.
In order to get a rough estimate of $g_{0,\mathrm{es \,SiO2}}$, we consider the silica as an isotropic material whose refractive index only depends on its density. Under this simplifying assumption,  the electrostrictive single photon coupling strength $g_{0,\mathrm{es \,SiO2}}$ is given by:
\begin{equation}
g_{0,\mathrm{es \,SiO2}}=\frac{\omega}{2}\frac{\int \gamma_e \, \tilde{\epsilon}_{v}\left(\vec{r}\right) E_p\left(\vec{r}\right) \, E_s\left(\vec{r}\right) \mathrm{d}^3 \vec{r} }{\sqrt{\int \varepsilon E_p\left(\vec{r}\right) \,\mathrm{d}^3 \vec{r}}\sqrt{\int \varepsilon E_s\left(\vec{r}\right) \,\mathrm{d}^3 \vec{r}}},
\label{Eq_g_0es2}
\end{equation}
where $E_p$ and $E_s$ respectively refer to the pump and Stokes fields, $\tilde{\epsilon}_v=\frac{\delta V_{zp}}{V}=-\frac{\delta\rho_{zp}}{\rho}$ is the zero-point volumetric strain caused by the Brillouin wave in the silica, and $\gamma_e=\left(\rho\, \frac{\partial\varepsilon}{\partial \rho}\right)_{\rho=\rho_0}=\left(\varepsilon-1\right)\left(\varepsilon+2\right)/3$ is the electrostrictive constant of the material 
\cite{boyd2003nonlinear}. (Thus $\gamma_e \, \tilde{\epsilon}_{v}$ corresponds to the zero-point permittivity fluctuations of the material due to the Brillouin wave). For backward scattering with identical azimuthal order pump and Stokes waves as is the case in our experiment, Eq.(\ref{Eq_g_0es2}) becomes:
\begin{equation}
g_{0,\mathrm{es \,SiO2}}=\frac{\omega}{2}\frac{\int \gamma_e \, \tilde{\epsilon}_{v}\left(\vec{r}\right) E^2\left(\vec{r}\right) \, \mathrm{d}^3 \vec{r} }{\int \varepsilon E^2\left(\vec{r}\right) \,\mathrm{d}^3 \vec{r}},
\label{Eq_g_0es1}
\end{equation}
We estimate $g_{0,\mathrm{es \,SiO2}}$ through the finite element simulations shown in Fig. \ref{fig:electrostrictive_g0}. Figure \ref{fig:electrostrictive_g0} (a) and (b) show the optical field distribution of the WGM used in the experiments, while (c) and (d) show the displacement and strain caused by an acoustic wave localized to the silica wedge with $m=2\times m_{\mathrm{opt}}$. Estimation of the single photon electrostrictive coupling rate through Eq. \ref{Eq_g_0es1} yields $g_{0,\mathrm{es\,SiO2}}=\sim17$ kHz. The radiation pressure contribution due to the moving boundary effect \cite{baker_photoelastic_2014} is much smaller, with $g_{0,\mathrm{rp\,SiO2}}=\sim 3$ Hz.
The electrostrictive $g_{0,\mathrm{es \,SiO2}}$ for the silica disk is comparable to the radiation pressure $g_{0,\mathrm{rp}}$ due to the moving boundary effect of the superfluid film.
The reason we do not observe Brillouin lasing from the silica disk is --beyond the approximatively three orders of magnitude higher acoustic damping rate for the GHz phonons in silica
\cite{lee_chemically_2012} compared to the superfluid sound wave--  that it not possible to get 3-fold resonant enhancement (pump+Stokes+acoustic wave) in the silica. Indeed, the Stokes beam would be strongly suppressed in our microresonator as there is a very low density of states for it 10 GHz away from the pump. Typically, the way to get efficient backward Brillouin scattering in a WGM microresonator is for the device's free spectral range (FSR) to match the Brillouin shift ($\sim$10 GHz). This requires a large device size, on the order of  6 mm diameter for silica \cite{Li2013,Li:17,lee_chemically_2012}. 

Miniaturization would enable large increases in the Brillouin interaction due the efficient co-localization of light and sound, but so far those efforts have been hampered by the fact that the usual energy and momentum matching criteria\footnote{Energy and momentum matching conditions in WGM resonators are given by $\omega_{\mathrm{pump}}=\omega_{\mathrm{Stokes}}+\ob$ and $m_B=m_{\mathrm{pump}} \pm m_{\mathrm{Stokes}}$, where + and - respectively correspond to the backward and forward scattering processes.} generally cannot be met in miniaturized devices with sparse optical spectra. 
However, because of the low speed of superfluid third-sound in our devices, this triple resonant enhancement is automatically satisfied independantly of device size. This  approach can be scaled all the way down to devices with $\lambda^3$ mode volumes \cite{gil-santos_high-frequency_2015}, enabling radiation-pressure $g_{0,\mathrm{rp}}/2\pi$ in excess of $10^5$ Hz with triply resonant interactions to be achieved.

\section{Optical strong coupling}
\label{sectionopticalstrongcoupling}

\subsection{Experimental measurement of strong optical coupling}

\begin{figure}
\includegraphics[width=\textwidth]{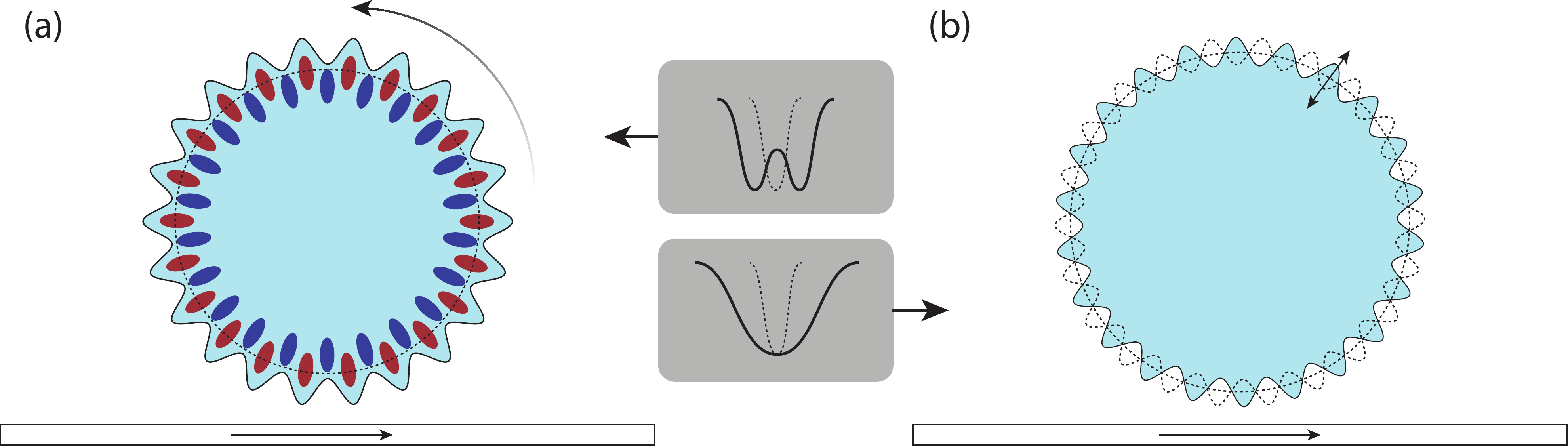}
\caption{Pump-probe measurement of the cavity optical response, for a travelling wave grating (a), and a stationary oscillating grating (b). While the superfluid film deformation is essentially out-of plane, its localized effect on the effective refractive index is represented here as a localized perturbation to the cavity radius. (a) The initially degenerate CCW and CW WGM resonances are split by the strong Brillouin-induced refractive index grating. The mode splitting can be described in the basis of quasi-stationary modes (red and blue WGMs) rotating at the speed of sound.} 
\label{Fig:mode_splitting}
\end{figure}

The optical spectrum of the cavity is measured with the following pump-probe setup, as illustrated in Figure 1 of the main text. A powerful ($\sim5 \,\mu$W) pump laser (NKT Koheras Adjustik fiber-laser) is brought onto resonance with the initially unsplit cavity WGM resonance. As the pump power is above threshold, this initiates spontaneous Brillouin lasing, as verified through the transmitted light power spectrum. This creates a travelling refractive index grating, as shown in Figure \ref{Fig:mode_splitting}(a). This grating travels counterclockwise at the speed of sound, with a frequency $\ob/2\pi\simeq 2\,c_3/\lambda_{\mathrm{light}}$. While the lasing is maintained, a low-power tunable diode laser (Yenista T100S-HP) is repeatedly swept across the optical resonance (100 Hz), using the laser's piezo-wavelength control. The transmitted optical photodetector signal is low-pass filtered (filter bandwidth = 20 kHz) and averaged (256 times) in order obtain the optical cavity spectrum, free from any modulation due to the Brillouin lasing process.

Due to the large refractive index grating generated by the Brillouin lasing process, the probe laser now  detects two distinct cavity resonances: a lower frequency resonance (red), where the light intensity is maximal under the peaks of the superfluid wave, and a higher frequency resonance (blue) ---spatially shifted by 
$\lambda_B/2$--- 
corresponding to maximal light intensity under the wave troughs, see Figure \ref{Fig:mode_splitting}(a). The magnitude of the measured splitting is a direct measure of the coupling rate $g_{opt}$ between pump and Stokes fields, and for $g_{opt} > \kappa$, strong mechanically mediated optical coupling between CCW and CW optical resonances is achieved.

Note that because the cavity decay rate $\kappa$ is much larger than the Brillouin frequency $\ob$ in this scheme, all intracavity photons sense an essentially static refractive index grating, as the grating motion is negligible over the $\sim 1/\kappa$ photon lifetime. Moreover, as the refractive index modulation is simply rotating around the device, the resonance frequencies of red and blue WGMs are essentially constant in time. As such, even though the probe laser sweep-rate is lower than the travelling grating frequency, the photodetector signal can be integrated without loss of information.

In contrast, for an `optomechanics-like' oscillating stationary refractive index grating (see Figure \ref{Fig:mode_splitting}(b)), the magnitude of the measured splitting would now be oscillating in time at frequency $2 \ob$. A time-averaged measurement would therefore reveal a broadened optical resonance, with no discernible splitting (see Fig. \ref{Fig:mode_splitting}(b) and e.g. Ref. \cite{winger_chip-scale_2011}).

\subsection{Theory}

Here we calculate the modified eigenfrequencies of a pair of optical cavity modes introduced by mechanically mediated scattering between them.

\begin{equation}
\hat H  = \hbar \omega_k a^\dagger_k a_k + \hbar \omega_{-k}a^\dagger_{k}a_{-k} - \hbar g_{0,\mathrm{rp}} \left (a^\dagger_k a_{-k} b_q + a_k a^\dagger_{-k} b_q^\dagger  \right ), 
\end{equation}
where $\omega_k$ and $\omega_{-k}$ are the bare eigen frequencies of the two optical modes, and the associated $a$ is the lowering operator of each mode. $g_{0,\mathrm{rp}}$ is the usual vacuum optomechanical coupling rate, and $b_q$ is the lowering operator for the co-propagating acoustic wave, where the counter-propagating wave is neglected. $k$ and $q$ are respectively the wave numbers of the optical modes and acoustic wave, with the signs in the front indicating directions.  The bare mechanical Hamiltonian has been neglected here, because we will instead just assume that the mechanical oscillator oscillates at some frequency and at some amplitude (the exact frequency will turn out not to matter, but should be expected to be very close to the bare mechanical resonance frequency).

This Hamiltonian models the Brillouin interaction of particular interest to this experiment, but also a broader class of "photon-phonon translator" type systems such as introduced by the Painter group~\cite{Safavi-Naeini2011}.

\subsection{Equations of motion}
\label{subsecequationsofmotiongopt}

From the Hamiltonian we directly obtain the equations of motion for the coupled optical cavity modes in the absence of dissipation. While dissipation can have the effect of shifting the eigen mode frequencies, for high quality optical cavities, as is the relevant case for this experiment, this is a negligible effect. Neglecting the dissipation allows a simpler calculation.

The equations of motion are found to be:
\begin{eqnarray}
\dot a_k &=& - i \omega_k a_k + i g_{0,\mathrm{rp}} a_{-k} b_q\\
\dot a_{-k} &=& -i \omega_{-k} a_{-k} + i g_{0,\mathrm{rp}} a_k b_q^\dagger
\end{eqnarray}

Then we treat the mechanical oscillator classically (we're interested in the light scattered due to its coherent oscillation rather than fluctuations) by substituting $b_q$ with $\beta_q e^{-i\ob t}$. $\beta_q$ is the amplitude of oscillation and $\ob$ can be thought of as the mechanical resonance frequency though in fact our results do not require this (of course, the further away from resonance the harder it will typically be to drive the oscillator to a given amplitude). We take $\beta_q$ to be real, without loss of generality. This just determines the phase of the mechanical oscillation. Defining the \emph{mechanical-amplitude boosted optomechanical coupling rate} $g_{\mathrm{opt}} = \beta_q \, g_{0,\mathrm{rp}}$, we then obtain
\begin{eqnarray}
\dot a_k &=& - i \omega_k a_k + i g_{\mathrm{opt}} e^{-i \ob t} a_{-k} \\
\dot a_{-k} &=& -i \omega_{-k} a_{-k} + i g_{\mathrm{opt}} e^{i \ob t} a_k.
\end{eqnarray}
The terms on the right are coherent coupling terms that act to hybridise the two optical modes. This results in a new pair of orthogonal eigenmodes with shifted frequencies. 

To determine the shifted eigenmode frequencies we postulate solutions of the following form:
\begin{eqnarray}
 a_k  &\rightarrow& \alpha_k e^{-i \omega t} \\
a_{-k} &\rightarrow&\alpha_{-k} e^{-i \omega t}
\end{eqnarray}
where $\omega$ is the oscillation frequency. Again we are neglecting the fluctuations in the fields, which do not alter the eigenfrequencies, with the $\alpha$'s representing the coherent amplitude of each field. We then find
\begin{eqnarray}
-i \omega a_k &=& - i \omega_k \alpha_k + i g_{\mathrm{opt}} e^{-i \ob t} \alpha_{-k} \\
-i \omega a_{-k} &=& -i \omega_{-k} \alpha_{-k} + i g_{\mathrm{opt}} e^{i \ob t} \alpha_k.
\end{eqnarray}
In matrix representation this can be written as
\begin{equation}
{\bf M} \cdot { \boldsymbol\alpha} = {\bf 0},
\end{equation}
where 
\[
{\bf M} =
\begin{bmatrix}
    \omega_k - \omega & -g_{\mathrm{opt}} e^{-i\ob t}  \\
    -g_{\mathrm{opt}} e^{i\ob t}  &  \omega_{-k} - \omega  \\
\end{bmatrix}
\]
and
\[
{\boldsymbol\alpha} =
\begin{bmatrix}
    \alpha_k   \\
    \alpha_{-k}   \\
\end{bmatrix}
\]
Non-trivial ($\boldsymbol\alpha \neq 0$) solutions to this matrix equation only exist when $\bf M$ is invertible and has a determinant equal to zero. This gives a condition on the frequency $\omega$:
\begin{equation}
|{\bf M}| = \left (\omega_k-\omega \right ) \left (\omega_{-k} - \omega \right) - g_{\mathrm{opt}}^2 = 0,
\end{equation}
which does not depend explicitly on the oscillation frequency $\ob$ of the mechanical element. Note, there is an implicit dependence, since achieving a large $\beta_q$ and therefore $g_{\mathrm{opt}}$ is easier for oscillation frequencies near the mechanical resonance frequency. Similarly, choosing appropriate bare optical frequencies can greatly enhance the ability of radiation pressure to drive the mechanical response. 

Solving this equation for $\omega$ gives two new shifted eigen frequencies $\omega_\pm$ given by
\begin{equation}
\omega_\pm = \bar \omega \pm \sqrt{\Delta^2/4 + g^2}, 
\end{equation}
where $\bar \omega = (\omega_k + \omega_{-k})/2$ is the average of the two bare resonance frequencies and $\Delta = \omega_k - \omega_{-k}$ is their difference. 

We can observe that when $g_{\mathrm{opt}} \gg \Delta$ the splitting between the resonances is given by $\delta = \omega_+-\omega_- = 2g_{\mathrm{opt}}$ as expected for strong coupling. However, in the reverse regime where $g \ll \Delta$, $\delta = \Delta +2 g_{\mathrm{opt}}^2/\Delta$. In this case, the first term ($\Delta$) is just the initial splitting of the resonances. The shift in splitting due to the mechanically-mediated coupling is $2 g_{\mathrm{opt}}^2/\Delta$, suppressed compared to the regime where  $g_{\mathrm{opt}} \gg \Delta$ by a factor of $\Delta/g_{\mathrm{opt}}$. In this superfluid Brillouin lasing experiment, the initial splitting of the two optical modes is zero, so the strong optical coupling is observed. In contrast, when the initial splitting of the two optical modes is engineered to be the Brillouin shift (typically $\sim 10$ GHz) in solid Brillouin experiments, it is not possible to obtain strong coupling of the two optical modes.

\subsection{Superfluid wave amplitude in the strong optical coupling regime}

As stated earlier, the superfluid acoustic wave-mediated optical coupling strength  takes the form:
\begin{equation}
\label{eq:OpticalCouping}
g_{opt} = \beta g_{0,\mathrm{rp}}.
\end{equation}  
Above the Brillouin lasing threshold, with 5.1 $\mu W$ pump power, the optical splitting in Fig. 3(b) of the main-text gives a mechanically mediated optical coupling rate of $g_{opt}/2\pi=187.4$ MHz. Combined with a single photon optomechanical coupling rate $g_{0,\mathrm{rp}}/2\pi=11$ kHz, obtained from fitting the experimental effective acoustic linewidth, this corresponds to a zero-point-motion normalised mechanical amplitude $\beta$ of $1.9\times 10^4$ (Eq. (\ref{eq:OpticalCouping})). 
This value is of comparable magnitude to that obtained by numerically solving the equations of motion above the lasing threshold (see section \ref{sectionBrillouinODE}). Indeed, as shown in Fig. 9(c), in those simulations the acoustic amplitude in the steady-state, with a slightly higher 5.1 $\mu W$ pump power,  is $\sim 4\times 10^4$ $x_{\mathrm{zpf}}$.  With a zero-point-motion amplitude $x_{\mathrm{zpf}}=9.5\times 10^{-15}$ m (see Table \ref{tableexperimentalparameters})
, these experimental and simulated amplitudes respectively correspond to a deformation of the superfluid interface of 
1.8 \AA $\,$  and  3.8 \AA.

\section*{References}
\footnotesize \renewcommand{\refname}{\vspace*{-30pt}}  